\documentclass[11pt]{article} 

\usepackage[utf8]{inputenc} 

\usepackage{geometry} 
\usepackage{graphicx} 

\usepackage[utf8]{inputenc} 
\usepackage{etex}
\usepackage{graphicx}
\usepackage{dcolumn}
\usepackage{amssymb,amsthm,amscd,amsbsy,array}
\usepackage{amstext}
\usepackage{amsmath,dsfont}
\usepackage{cancel}
\usepackage{soul} 
\usepackage{graphics,xcolor}
\usepackage{bm}
\usepackage{booktabs} 
\usepackage{array} 
\usepackage{paralist} 
\usepackage{verbatim} 
\usepackage{subfig} 

\numberwithin{equation}{section}

\let\ssection=\section
\renewcommand{\section}{\setcounter{equation}{0}\ssection}
\textheight=
250mm 
\newcommand{\half}{{\frac{1}{2}}}
\def\2{{\half}}

\def\bnabla{\mbox{\boldmath$\nabla$}}

\def\bnabla{{\bm{\nabla}}}

\def\bx{{\bm{x}}}

\def\bX{{\bm{X}}}

\def\beq{\begin{equation}}
\def\eeq{\end{equation}}
\def\beqa{\begin{eqnarray}}
\def\eeqa{\end{eqnarray}}

\def\barray{\left(\begin{array}}
\def\earray{\end{array}\right)}
\def\barraynb{\begin{array}}
\def\earraynb{\end{array}}

\def\smallover#1/#2{\hbox{$\textstyle\frac{#1}{#2}$}} %

\newcommand{\Tr}{\mathrm{Tr}}

\usepackage{color}

\newcommand{\gb}{\colorbox{green}}

\def\benu{\begin{enumerate}}
\def\eenu{\end{enumerate}}

\def\?{{\;\gb{\,?\,} \;}}

\usepackage{fancyhdr} 
\usepackage{sectsty}
\allsectionsfont{\sffamily\mdseries\upshape} 

\usepackage[nottoc,notlof,notlot]{tocbibind} 
\usepackage[titles,subfigure]{tocloft} 

\title{Various disguises of the Pais-Uhlenbeck oscillator}
\author{ Mahmut Elbistan$^{1}$\footnote{mailto: elbistan@itu.edu.tr} \  and Krzysztof Andrzejewski$^2$\footnote{mailto: krzysztof.andrzejewski@uni.lodz.pl} }

\date{} 

\begin{document}

\maketitle

1) Physics Department, 
Bo\u{g}azi\c{c}i University, 
34342 Bebek / Istanbul, Turkey

2) Faculty of Physics and Applied Informatics,
University of Lodz, Pomorska 149/153, 90-236,  Lodz, Poland


\begin{abstract}
Beginning with a simple set of planar equations, we discuss novel realizations of the Pais-Uhlenbeck oscillator in various contexts. First, due to the bi-Hamiltonian character of this  model, we develop a Hamiltonian approach for  the Eisenhart-Duval lift of the related dynamics. We apply this approach to the previously worked example of a circularly polarized periodic gravitational wave. Then, we present our further results. Firstly, we show that the transverse dynamics of the Lukash plane wave and a complete gravitational wave pulse  can also lead to the Pais-Uhlenbeck oscillator. We express the related Carroll Killing vectors in terms of the  Pais-Uhlenbeck frequencies and derive extra integrals of motion from the conformal Newton-Hooke symmetry.
In addition, we find that the $3+1$ dimensional Penning trap can be canonically mapped to the 6th order Pais-Uhlenbeck oscillator. We also carry the problem to the non-commutative plane. Lastly, we discuss other examples like the motion of a charged particle under electromagnetic field created with double copy.

\end{abstract}

\pagebreak

\tableofcontents

\newpage


\section{Introduction}

It has been quite a time since Pais and Uhlenbeck \cite{PU} invented their $1+1$-dimensional non-relativistic (NR) mechanical model, now called the Pais-Uhlenbeck (PU) oscillator, as a prototype for higher order theories (see e.g. \cite{smilga} for a review). Namely, it is  defined by the  following   Lagrangian 
\beq
\label{lpu}
L= -\frac 12 {x} \prod_{k=1}^{n}\left(\frac{d^2}{dt^2} +\lambda^2_k\right) {x}.
\eeq
One of the crucial  observations \cite{PU}   is that  the Hamiltonian  of the $2n$th order PU oscillator can be written as a combination of $n$ decoupled harmonic oscillators with an alternating sign 
\begin{equation} 
\label{PUHOs}
H = \frac{1}{2} \sum_{k=1}^n (-1)^{k+1} (p_k^2 + \lambda_k^2 x_k^2).
\end{equation}
As a feature of higher derivative theories, the energy (\ref{PUHOs}) is not positive definite. Ostrogradski's Hamiltonian for the PU oscillator necessarily has a linear term in one of the momenta\footnote{See eqn. (\ref{PUHorigin}) for $n=2$.} and the associated Lagrangian begins with a square of $n$th derivative of the position. 

Since it is an interesting  topic,  the PU oscillator has been the subject of further studies. For instance, its quantization was discussed \cite{Bender:2007wu}, interactions were added \cite{Pavsic:2013noa, Pavsic:2013mja, Pavsic:2016ykq}, odd order generalization was made \cite{Masterov:2016jft} and lastly damped case was worked out \cite{Masterov:2022fai}. Its relation to curl forces and to gyroscopic systems were also noted recently in \cite{Guha:2020kmk} and \cite{Comelli:2022evf}, respectively.

The goal of this paper is to reveal another aspect of the PU oscillator, namely its realization in various and seemingly different contexts like gravitational waves and NR ion traps. Moreover, following \cite{Hill2, Zhang:2012cr} we give a formulation of the PU oscillator in non-commuting coordinates.  We further relate the motion of a charged particle in the electromagnetic background created via double copy \cite{Andr1} to the $4$th order PU oscillator. So, in spite of being a higher derivative theory, PU oscillator appears in the dynamics of several interesting physical systems with second order equations of motion.

For our aim, an important clue comes from the symmetries of the PU oscillator. Let us note that the symmetries of the $4$th order PU oscillator span 2 copies of  the Heisenberg algebra for generic $\lambda_i$s \cite{Andrzejewski:2014zia}. In the case of odd frequencies the algebra becomes richer: It is the $\ell = \frac{2n-1}{2}$ conformal Newton-Hooke (NH) algebra \cite{Andrzejewski:2014rza} which is equivalent to the Galilei algebra up to a redefinition. There has been a renewed interest for this group in the field of fluid dynamics \cite{Galajinsky:2022ywb, Galajinsky:2022rfu, Snegirev:2023obh}.    

Recently in \cite{Elbistan:2022umq} it was shown that the $1+1$-dimensional 4th order PU oscillator can be related to a $1+3$ dimensional circularly polarized periodic gravitational wave (CPP GW) via the Eisenhart-Duval (ED) lift a.k.a. Bargmann framework \cite{Eisenhart, DuvalBargmann, DGH,CGGH}. The ED lift provides a correspondence between a certain class of Riemannian manifolds called the Bargmann manifolds and lower dimensional NR systems. Basically, the transverse part of the geodesic equations of the former coincide with the Newtonian equations of the latter. The correspondence shown in \cite{Elbistan:2022umq} is surprising because it is generically assumed that for a $1+3$ dimensional plane wave, the associated NR theory is $2+1$ dimensional. There is no contradiction though as it is the phase space dimension which counts and the phase space of a $1+1$ dimensional PU oscillator is $4$ dimensional just as a $2+1$ dimensional NR theory. 

In the case of the plane gravitational  waves,  the transverse part of  the geodesic equations define anisotropic, time-dependent and coupled oscillators in the form of Sturm-Liouville (SL) problem \cite{Zhang:2018gzn} which is very difficult to solve. However, the situation changes if there exists an extra symmetry. In the former example of the CPP GW, apart from the generic $5$-parameter Carroll symmetry of the plane wave \cite{Duval:2017els} and the homothety \cite{phonebook}, there exists an extra screw isometry that led us to new coordinates where equations become time-independent and easy to solve. Then, the Carroll symmetries were projected and became NH charges for the corresponding PU oscillator \cite{Elbistan:2022umq}. 

Apart from the CPP GW, there exists a second vacuum plane wave solution called the Lukash plane wave \cite{phonebook} which is endowed with an extra isometry. Its geodesic motion was studied in \cite{Zhang:2021lrw}. There is also a geodesically complete vacuum plane wave profile studied in \cite{Andr1, Andr2} which has an extra conformal symmetry. In both examples, using those extra symmetries, transverse part of geodesic motion can be cast into the form
\begin{subequations}
\label{gcosc}
\begin{align}
x'' - \omega y' - \Omega_+^2 x&= 0, \\
y'' + \omega x' - \Omega_-^2 y&=0.
\end{align}
\end{subequations}
Here, for the first time, we will relate those transverse motions in the Lukash and the complete plane waves to a $1+1$ dimensional $4$th order PU oscillator.

After ending our discussion about plane waves, next we consider lower dimensional NR physics. In that context, our first example will be the $3+1$-dimensional Penning trap \cite{Dehmelt}, an experimental device to store molecular ions. The ED lift of the Penning trap and the related geodesic motion was previously studied in \cite{IONGW}. Here, we will show that the planar dynamics of the Penning trap is described by (\ref{gcosc}) and with the addition of the third direction it can be surprisingly mapped into a $1+1$ dimensional but $6$th order PU oscillator. The physical parameters of the Penning trap, namely modified cyclotron, magnetron and axial frequencies turn out be frequencies $\lambda_1, \lambda_2, \lambda_3$ of the PU oscillator (\ref{PUHOs}) in a nice way.

It is known that in $2+1$ dimensions the Galilei group admits a second central extension (see \cite{Horvathy:2010wv} and references therein). This situation can be realized in condensed matter systems where the momentum dependent Berry phases come into play \cite{Niu}. We work out equations (\ref{gcosc}) in the  non-commutative plane by simply postulating a non-commutativity parameter $\theta$, see (\ref{symcosctheta}). As a result, NH symmetry is enhanced with a second central extension and we show that even this exotic model can be mapped to a $4$th order $1+1$ dimensional PU oscillator with $\theta$-dependent frequencies. This new result may yield potential applications of the PU oscillator in condensed matter theory.   

Lastly, we discuss the interrelation between the PU oscillator and the motion of a charged particle in electromagnetic fields created by double copy \cite{ilderton, Andr1}, Hill's equations \cite{Hill}, Lagrange points \cite{Bialynicki-Birula:1994bie} and gravitational trapping \cite{Bialynicki-Birula:2018nnk}. 

The outline of the paper is as follows: At Section \ref{trapPU}, we begin with our generic planar trapping problem and link it to the PU oscillator.  At Section \ref{planewaves}, we discuss the related ED lifted plane wave metrics and their symmetries. We also develop a Hamiltonian approach to avoid the ambiguity related to the bi-Hamiltonian  aspect of the PU model.  As a bonus, we find the canonical transformation between the underlying NR systems of the  Brinkmann and  Baldwin, Jeffery and Rosen (BJR) metrics.  At Section \ref{CPProt}, we revisit  the CPP GW case and write explicitly its Carroll Killing vectors and  additional symmetries originating from the NH symmetry of the PU oscillator. Sections \ref{Lukash} and \ref{complete} are devoted to the Lukash plane wave and the complete pulse example and their relation to the PU oscillator, respectively. At section \ref{Penning}, we show that the Penning trap can be mapped to a $n=3$ PU oscillator.  At section \ref{exotic}, the coordinates are promoted to be non-commuting, then   the associated exotic problem and its symmetries are discussed. We mention other related examples at Section \ref{other}. Finally, at Section \ref{discussion} we summarize our results and comment on future problems.


\section{Trapping and the PU oscillator}
\label{trapPU}


\subsection{Setting of the problem}
Let us consider  the Lagrangian 
\begin{equation}
\label{Lgcosc}
L = \frac{1}{2} M (x'^2 + y'^2) - \frac{M \omega}{2} (y x' - x y') \mp \frac{M}{2} (\Omega_+^2 x^2 + \Omega_-^2 y^2),
\end{equation}
where $M$ is a non-zero parameter. The corresponding Hamiltonian is of the form 
\begin{equation}
\label{Hgcosc}	
H = \frac{p_x ^2+p_y^2}{2M} +\frac{\omega}{2}(y p_x-xp_y)+\frac{M\omega^2}{8}(x^2+y^2)  \pm \frac{M}{2} (\Omega_+^2 x^2 + \Omega_-^2 y^2).
\end{equation}
Note that by the   replacement    $H\rightarrow -H, \omega \rightarrow -\omega $ we can restrict to $M>0$; moreover using the  coordinate change $x \leftrightarrow y$   we may choose $\omega > 0$.
The Hamiltonian \eqref{Hgcosc}  for   $+$ sign   describes a charged particle in a static homogeneous magnetic field  (in the symmetric gauge) and an anisotropic  harmonic potential. 
Because of the loss of symmetry caused by the anisotropic  term  we cannot reduce  it  to  a one-dimensional problem in the standard way. Despite this fact,  there is  a canonical transformation  leading to the  sum of  two independent  harmonic oscillators (in particular, the energy  is positive-definite), see \cite{Pl1, Pl2, BergHolz, DSC}.  

However, for our  further purposes  more interesting will be  the second case, i.e. with  the inverted anisotropic potential (minus sign in \eqref{Hgcosc}).   
To get  some insight into this choice  of  sign, first, let us analyse the corresponding equations of motion (\ref{gcosc})
where derivation $\{'\}$ is w.r.t. to NR time. We assume no prior relations between $\omega$ and $ \Omega_\pm$ so that our findings can be applied to various cases.  
Note that without  the Coriolis force i.e., $\omega=0$, the motions will be uncoupled and they will be described by hyperbolic functions\footnote{It is not excluded that those motions can be described by trigonometric functions when 
$\Omega_\pm^2 < 0$. In that case one should be careful about reality/hermiticity of the related Hamiltonian. Here we consider the case $\Omega_\pm >0$.}. It is this Coriolis force or magnetic field that makes trapping possible (within a suitable parameter range for $\omega, \Omega_\pm$). Similar logic applies to a Penning trap \cite{Dehmelt} where motions are stabilized via a constant magnetic field. Alternatively, one can think of equations (\ref{gcosc}) as a generalized version of planar Hill's equations for the Earth-Moon-Sun system \cite{Hill}.

Eqs. \eqref{gcosc} can be transformed into a set of two  decoupled linear oscillators; so the corresponding  Hamiltonian should be   a   combination of  two oscillator Hamiltonians.  Since we started  with inverted anisotropic oscillators, the Hamiltonian is not positively defined. Thus, instead of the sum, for specific values of parameters we may obtain  the difference of two oscillators and  consequently the PU model (\ref{PUHOs}). Before we  perform  a detailed analysis at the Hamiltonian level confirming this conjecture; first, let us have a look at this problem more directly.
Deriving and substituting equations in each other we get a decoupled but the 4th order equation either for $x$ or $y$ as
\begin{equation}
\label{4thPU}
x'''' + (\omega^2 - \Omega_+^2 - \Omega_-^2) x'' + \Omega_+^2 \Omega_-^2 x =0.
\end{equation}
Comparing eq.   \eqref{4thPU} with the equation of motion following from   \eqref{lpu} (for $n=2$)
\begin{equation}
x'''' +(\lambda_1^2+\lambda_2^2)x''+\lambda_1^2\lambda_2^2x=0,
\end{equation} 
we conclude that to obtain the PU model  $\lambda_{1,2}$ should satisfy the following conditions
\begin{equation}
\label{LOcond}
\lambda_1^2 \lambda_2^2 = \Omega_+^2 \Omega_-^2,    \quad   \lambda_1^2 + \lambda_2^2 = \omega^2 - \Omega_+^2 -\Omega_-^2.
\end{equation}
In the particular case of $\omega^2 = 2 (\Omega_+^2 + \Omega_-^2)$ the equation (\ref{4thPU}) automatically takes the  PU form  \cite{Elbistan:2022umq}.
These equations are symmetric under $\lambda_1 \leftrightarrow \lambda_2$;  for $\lambda_2^2$ we found the algebraic relation
\begin{equation}
\lambda_2^4 - (\omega^2 - \Omega_+^2 - \Omega_-^2)\lambda_2^2 + \Omega_+^2 \Omega_-^2 =0, 
\end{equation} 
that yields two roots  
\begin{equation}
\label{e10}
\lambda_2^2 = \frac{(\omega^2 - \Omega_+^2 - \Omega_-^2) \pm \sqrt{\Delta}}{2}, \quad \quad    \Delta =  (\omega^2 - \Omega_+^2 - \Omega_-^2)^2 - 4 \Omega_+^2 \Omega_-^2.  
\end{equation}
We make the choice 
\begin{equation}
\label{lambdas}
\lambda_1^2 =  \frac{(\omega^2 - \Omega_+^2 - \Omega_-^2) + \sqrt{\Delta}}{2}, \qquad  \lambda_2^2 =  \frac{(\omega^2 - \Omega_+^2 - \Omega_-^2) - \sqrt{\Delta}}{2}. 
\end{equation}
However, we should  consider only real and positive $\lambda^2$'s,    thus some additional  relations on $\Omega_\pm^2$ and $\omega$ are needed. First,  $|\omega^2-\Omega_+^2-\Omega_-^2|> 2\Omega_+\Omega_-$; however,  for  $\omega^2-\Omega_+^2-\Omega_-^2<0$ we have $\lambda_2^2<0$ thus we assume  $\omega^2-\Omega_+^2-\Omega_-^2>0$.   In view of this  for   $\omega> \Omega_++\Omega_-$  we obtain  real and positive ${\lambda^2}'s$. 
\par
Although the above  intuitive  approach is useful to obtain  necessary  conditions for   the PU model, it may be misleading.  Had we chosen the first case with  plus sign in (\ref{Hgcosc}), again we would have ended up with a $4$th order equation analogous to (\ref{4thPU}). However,   in this  case the resulting theory is the sum  of two harmonic oscillators \cite{Zhang:2012cr, BergHolz, DSC}.  Such a situation   is related  to the bi-Hamiltonian  property of the PU model \cite{BolonekKosinski}.   Therefore, a more rigorous and explicit link to PU oscillator at the Hamiltonian level will be presented in the next section. 

\par 
Based on the analysis made in \cite{Pavsic:2013noa, BolonekKosinski}, we note that there might be other cases where our results can be straightforwardly extended. For instance, when one of the frequencies, say $\Omega_- $, is vanishing, the solution (\ref{lambdas}) again works with  $\lambda_2 = 0$ and $\lambda_1^2 = \omega^2 - \Omega_+^2 $. This case will be separately handled at section \ref{Hill} when we discuss Hill's equations \cite{Hill}. Let us also mention that when $\Delta < 0$, $\lambda_{1,2}^2$ are complex conjugate to each other.


\subsection{Explicit link to PU oscillator}

The relation between  second order coupled equations (\ref{gcosc}) and the PU model has  been recently discussed in various ways and contexts, including curl forces and optics \cite{Guha:2020kmk},  Coriolis  force and CCP GW \cite{Elbistan:2022umq},  gyroscopic systems and dark energy \cite{Comelli:2022evf}. Since, for our further considerations the Hamiltonian  \eqref{Hgcosc} with the minus sign will be the starting point,  we need to generalize the approach in \cite{Elbistan:2022umq}.  To this end let us introduce the kinematical momenta $\Pi_i = (  \Pi_x, \Pi_y)$ as
\begin{equation}
\label{kinemom}
\Pi_i = p_i + \frac{M \omega}{2} \epsilon_{ij} x^j, 
\end{equation}
and write the associated symplectic structure in terms of that non-canonical coordinates $(x, y, \Pi_x, \Pi_y )$ 
\begin{subequations}
\label{symgcosc}
\begin{align}
H &= \frac{\Pi_x^2}{2M} + \frac{\Pi_y^2}{2M} - \frac{M}{2} (\Omega_+^2 x^2 + \Omega_-^2 y^2), \\
\sigma &= d \Pi_i \wedge dx^i + M \omega dx \wedge dy .
\end{align}
\end{subequations}
The symplectic structure (\ref{symgcosc}) directly yields equations (\ref{gcosc}).

Our goal is to apply the method of chiral decomposition \cite{Pl1, Pl2} and split the coupled theory (\ref{symgcosc}) above. This amounts to trade the  phase space coordinates $(x, y, \Pi_x, \Pi_y )$  with the new ones $(X_+^{1,2},  X_-^{1,2})$  
\begin{subequations}
\label{cdcoef}
\begin{align}
\Pi_x &= \alpha_+  X_+^2 + \alpha_-  X_-^2,   \quad  \Pi_y = - \beta_+ X_+^1 - \beta_- X_-^1, \\  
x &= X_+^1 + X_-^1, \quad \quad  \quad y = X_+^2 + X_-^2 ,
\end{align}
\end{subequations}
and solve for the following equations
\begin{subequations}
\label{coefcdgcosc}
\begin{align}
\beta_+ \beta_- &= M^2 \Omega_+^2,    \quad  \quad    \alpha_+ \alpha_- = M^2 \Omega_-^2, \\
\alpha_+ + \beta_- &= M\omega,  \quad  \quad   \alpha_- + \beta_+ = M \omega .
\end{align}
\end{subequations}

Omitting details, below we present our solutions for the constants $\beta_\pm, \alpha_\pm$ in terms of diagonalization frequencies (the  PU frequencies) $\lambda_{1,2}$ (\ref{lambdas}) as 
\begin{subequations}
\label{cdgcosc}
\begin{align}
\beta_- &= \frac{M}{\omega} (\lambda_1^2 + \Omega_+^2) , \quad \quad  \beta_+ = \frac{M\omega \Omega_+^2}{\lambda_1^2 + \Omega_+^2}, \\
\alpha_+&= \frac{M}{\omega} (\lambda_2^2 + \Omega_-^2) , \quad \quad \alpha_- = \frac{M\omega \Omega_-^2}{\lambda_2^2 + \Omega_-^2} . 
\end{align}
\end{subequations}
Substituting them into eqs.  (\ref{symgcosc}), we obtain a symplectic structure which is decomposed into $+$ and $-$ sectors
\begin{subequations}
\label{symdegeneric}
\begin{align}
H &= - \frac{M (\lambda_1^2 - \lambda_2^2)}{2} \left( \frac{\lambda_2^2}{(\lambda_2^2 + \Omega_-^2)} (X_+^1)^2 + \frac{\lambda_2^2 + \Omega_-^2}{\omega^2} (X_+^2)^2  -  \frac{\lambda_1^2}{(\lambda_1^2 + \Omega_+^2)} (X_-^2)^2 -   \frac{\lambda_1^2 + \Omega_+^2}{\omega^2} (X_-^1)^2      \right), \\
\sigma &= \frac{M (\lambda_1^2 - \lambda_2^2)}{\omega} (dX_+^1 \wedge dX_+^2 - dX_-^1 \wedge dX_-^2) . 
\end{align}
\end{subequations}
Let us remind that in the above formulae, we assume  that $\lambda_i^2$s (\ref{lambdas}) are real and positive (see the previous section).  
Although (\ref{symdegeneric}) looks a bit complicated as it stands, the related Euler-Lagrange equations become the ones for  the  PU oscillator with  frequencies $\lambda_{1,2}$ (\ref{lambdas})  \footnote{At this point, one can easily find generic solutions for $x, \ y$ (\ref{cdcoef}b) to show that they both solve for the PU equation (\ref{4thPU}) and (\ref{gcosc}). }
\begin{equation}
(X_+^{1,2})'' + \lambda_2^2 X_+^{1,2} = 0, \quad   (X_-^{1,2})'' + \lambda_1^2 X_-^{1,2} = 0.
\end{equation}
When we pass to Darboux coordinates  by means of the following simple redefinition
\begin{subequations}
\begin{align}
 \sqrt{\frac{M (\lambda_1^2 - \lambda_2^2)(\lambda_1^2 + \Omega_+^2)}{\omega^2}} X_-^1 &= -p_1,  \quad \quad   \sqrt{\frac{M(\lambda_1^2 - \lambda_2^2)}{\lambda_1^2 + \Omega_+^2}} X_-^2 = x^1, \\
 \sqrt{\frac{M (\lambda_1^2 - \lambda_2^2)(\lambda_2^2 + \Omega_-^2)}{\omega^2}} X_+^2 &= - p_2,  \quad  \quad \sqrt{\frac{M(\lambda_1^2 - \lambda_2^2)}{\lambda_2^2 + \Omega_-^2}} X_+^1 = x^2 ,
\end{align}
\end{subequations}
we see that the Hamiltonian consists of two harmonic oscillators, surprisingly with a relative minus sign between them:  
\begin{equation}
\label{Hopm}
H = H_1 - H_2, \qquad   H_i = \frac{1}{2}\left( p_i^2 + \lambda_i^2 (x^i)^2  \right),   
\end{equation}
cf. (\ref{PUHOs}). Finally,  there exists a final canonical transformation \cite{smilga}
\begin{subequations}
\label{scanon}
\begin{align}
x^1 &= \frac{ p_x +  \lambda_1^2 v }{\lambda_1 \sqrt{(\lambda_1^2 - \lambda_2^2)}},  \quad   p_1 = \frac{\lambda_1 (p_v + \lambda_2^2 x)}{ \sqrt{(\lambda_1^2 - \lambda_2^2)}}, \\
x^2 & = \frac{p_v + \lambda_1^2 x }{\sqrt{(\lambda_1^2 - \lambda_2^2)}} ,   \quad p_2  = \frac{p_x + \lambda_2^2 v}{ \sqrt{(\lambda_1^2 - \lambda_2^2)}},
\end{align}
\end{subequations}
which maps (\ref{Hopm}) to Ostragadski's Hamiltonian of the PU oscillator, 
\beq
\label{PUHorigin}
H = H_{PU} = p_x v + \frac{p_v^2}{2} + \frac{(\lambda_1^2 + \lambda_2^2)v^2}{2}  -  \frac{\lambda_1^2 \lambda_2^2 x^2}{2} .  
\eeq 

To sum up,  the  Hamiltonian \eqref{Hgcosc},  for  the case of  minus sign, can lead to the PU oscillator with $\lambda^2$s being positive.    
The question is in what context such a peculiar choice of potential and, in consequence, a possible realization of the PU dynamics appear. This is  the  main aim of our further considerations. 
In particular, we  will show that for some  gravitational and electromagnetic fields and for a Penning trap such a  situation  occurs.


\section{Plane waves, ED lift and Hamiltonian formalism}
\label{planewaves}

In view of the above considerations the crucial question is whether we can find some physically interesting models with the Hamiltonian \eqref{symgcosc}. The hint to this problem is the so-called ED lift.  Namely, it turns out that for some plane wave space-times the transverse part of the geodesic equations decouple and take the form of the 2-dimensional (time-dependent) Newton's equations. Conversely, some classical equations (in particular eq. \eqref{gcosc}) can be embedded into higher dimensional relativistic geodesics. Such an approach is known as the ED lift and has been studied in various places and contexts (see e.g., \cite{Cariglia:2014ysa, Cariglia:2015bla} for a review).  
However, there is a subtle point related to the fact the PU oscillator is a bi-Hamiltonian system  \cite{BolonekKosinski} what, in turn, involves  to use the Hamiltonian formalism  rather than  equations of motion. We will address to this issue at Section \ref{altH}. Firstly, we will discuss the lift of (\ref{gcosc}) and its symmetry aspects.


\subsection{ED lift and symmetries}
\label{EDlifts}
 
When the NR system \eqref{symgcosc} is  ED lifted, we get the following plane-wave metric 
\begin{equation} 
\label{Liftgcosc}
ds^2 =g_{\mu\nu}dx^\mu dx^\nu = dx^2 + dy^2 + 2 du dv + \omega(x dy - y dx) du + (\Omega_+^2 x^2 + \Omega_-^2 y^2) du^2,  
\end{equation}
in rotating Brinkmann coordinates $x^\mu = (x, y, u, v)$.
Conversely, we can use   the null geodesic condition 
\begin{equation}
\label{Hgeo4d}
H_{geo} = \frac{g^{\mu\nu}p_\mu p_\nu}{2} \equiv 0,   \qquad  p_\mu = (p_i, p_u, p_v), 
\end{equation} 
to  recover the underlying NR dynamics (\ref{gcosc}) where $u$ plays the role of NR time and its conjugate momentum $p_u$ becomes  $-H$ (\ref{symgcosc}). Coordinate $v$ turns out to be the associated NR action and the conserved $p_v$ is identified with the NR mass $M$.

The plane waves have a 5-parameter isometry group which is isomorphic to the  $2+1$ dimensional Carroll group with broken rotations \cite{Duval:2017els} (or 2 copies of Heisenberg algebra). Finding Killing vectors amounts to deal with a matrix SL equation and only a few exact solutions are known.  

Having already solved the underlying NR motion, we computed the Killing vectors of the metric (\ref{Liftgcosc}) in terms of the  related PU oscillator frequencies $\lambda_{1,2}$ as\footnote{Note that we assume $\lambda_{1,2}^2 > 0$, $\sqrt{\Delta} = \lambda_1^2 - \lambda_2^2$ and $\Omega_\pm > 0$ while deriving (\ref{CarrollrotBgen}). Situation changes for Hill's equations where $\Omega_- = \lambda_2 = 0$, see Section \ref{Hill}. }
\begin{subequations}
\label{CarrollrotBgen}
\begin{align}
Y_1^c &= \frac{1}{2(\lambda_2^2 + \Omega_-^2)} \Big(  \omega(\lambda_2^2 - \Omega_-^2)y \cos\lambda_2 u - \lambda_2 (\Omega_+^2 -\Omega_-^2 + \sqrt{\Delta})x\sin\lambda_2 u \Big)\partial_v + \cos\lambda_2 u \partial_x - \frac{\omega \lambda_2}{\lambda_2^2 + \Omega_-^2}\sin\lambda_2u\partial_y , \\
Y_2^c &= \frac{1}{2(\lambda_1^2 + \Omega_+^2)} \Big(-\omega(\lambda_1^2 - \Omega_+^2)x \cos\lambda_1 u + \lambda_1 (\Omega_+^2 -\Omega_-^2 + \sqrt{\Delta})y\sin\lambda_1 u \Big)\partial_v + \frac{\omega \lambda_1}{\lambda_1^2 + \Omega_+^2}\sin\lambda_1u\partial_x +  \cos\lambda_1 u \partial_y ,  \\
Y_1^b &= -\frac{\omega}{\sqrt{\Delta}} \left[ \Big(\frac{(\Omega_+^2 - \Omega_-^2 +\sqrt{\Delta})}{2 \omega} x \cos\lambda_2 u + \frac{(\lambda_2^2- \Omega_-^2)}{2\lambda_2} y\sin\lambda_2u \Big)\partial_v + \frac{(\lambda_2^2 + \Omega_-^2)}{\omega\lambda_2}\sin\lambda_2u\partial_x + \cos\lambda_2u\partial_y \right] ,\\ 
Y_2^b &= -\frac{\omega}{\sqrt{\Delta}} \left[ \Big(\frac{(\Omega_+^2 - \Omega_-^2 +\sqrt{\Delta})}{2 \omega} y \cos\lambda_1 u + \frac{(\lambda_1^2- \Omega_+^2)}{2\lambda_1} x\sin\lambda_1u \Big)\partial_v  + \cos\lambda_1u\partial_x  - \frac{(\lambda_1^2 + \Omega_+^2)}{\omega\lambda_1}\sin\lambda_1u\partial_y  \right].
\end{align}
\end{subequations}
The Carroll translations $Y_i^c$ and Carroll boosts $Y_i^b$ are completed with the  $v-$translation  $Y^v = \partial_v$ to satisfy the aforementioned Carroll algebra 
\begin{equation}
\label{Carrollplane}
[ Y_i^c, Y_j^b] = - \delta_{ij}\partial_v.
\end{equation}
Apart from the generic homothetic vector field $Y_h = 2 v\partial_v + x^i \partial _i$, our metric (\ref{Liftgcosc}) has an extra isometry which is a $u$-translation $\partial_u$.
Thus we come to the  conclusion that if there exists a plane wave whose underlying NR system can be mapped to the  $1+1$ dimensional PU oscillator, it should be endowed with an additional symmetry\footnote{Extra isometry $\partial_u$ corresponds to the conserved Hamiltonian of the PU oscillator (\ref{PUHorigin}) after null projection \cite{Elbistan:2022umq}.}. 

Coming to possible cases, there are only 2 exact plane gravitational waves endowed with an extra isometry \cite{phonebook}. First one is the CPP GW whose relation to the PU oscillator through  the ED lift was worked out in \cite{Elbistan:2022umq}. 
 The second vacuum example with an extra isometry is the Lukash plane wave and its geodesic motion was studied in \cite{Zhang:2021lrw}. At Section \ref{Lukash}, we will show that it can be also related to the PU oscillator in a non-trivial way.
Moreover, there is a third example of polarized  plane gravitational waves with an extra conformal symmetry \cite{Andr2}. We will study this example at Section \ref{complete}  and will  show that its underlying theory can be described by a 1-dimensional PU oscillator.

Killing vectors (\ref{CarrollrotBgen}) will help us to spell out the symmetries of various systems that we link to the PU oscillator in the following sections. For instance, using them we write down the isometries of CPP GW (\ref{Carrollrot}) that become the NH symmetry of the PU oscillator when projected.  They also allow us to derive conserved charges of the Penning trap (\ref{CarrollPenning}) which will be linked to a $1+1$ dimensional but  the $6$th order PU oscillator. 

Finally, let us make the following change
\begin{equation}
\label{e9} 
\bx=R^{-1}(U)\bX, \quad u= U, \quad v = V ,
\qquad \qquad
R(U)=\left(
\begin{array}{cc}
\cos(\omega U/2) &-\sin(\omega U/2)\\
\sin(\omega U/2) & \cos(\omega U/2)
\end{array}
\right),
\end{equation}
and express (\ref{Liftgcosc}) in the standard Brinkmann coordinates $X^\mu = (X^1, X^2, U, V)$
\begin{eqnarray}
\label{Liftgcoscunr}
ds^2 &=& d\bm{X}^2 + 2 dU dV + K_{ij}(U) X^i X^j dU^2, \\
K_{11} &=& \Omega_+^2 \cos^2(\omega U/2) + \Omega_-^2 \sin^2(\omega U/2) - \frac{\omega^2}{4}, \nonumber\\
K_{22} &=& \Omega_+^2 \sin^2(\omega U/2) + \Omega_-^2 \cos^2(\omega U/2) - \frac{\omega^2}{4}, \nonumber\\
K_{12} &=& \sin(\omega U/2) \cos(\omega U /2) (\Omega_+^2 - \Omega_-^2)  \nonumber.
\end{eqnarray}
The trace of the $2\times 2$ $K$ matrix is $ \Tr{(K)} = \Omega_+^2 + \Omega_-^2 - \frac{\omega^2}{2}$. When $\Tr{(K)} = 0$, metric (\ref{Liftgcoscunr}) is a vacuum solution of Einstein's equations and it is called as an exact plane gravitational wave. CPP and Lukash plane waves belong to this class. So, depending on the parameters $\omega, \Omega_\pm$ our metric at hand can be a vacuum solution or not.

In the standard Brinkmann coordinates (\ref{Liftgcoscunr}), the extra symmetry becomes the screw isometry
\begin{equation}
\label{screwforgen}
Y_s = \partial_U + \frac{\omega}{2} \epsilon_{ij} X^i \partial_j, 
\end{equation}
which is a combination of a $U$ translation and a half-angle rotation on the plane. Moreover, in that coordinate system the metric is in the Kerr-Schild form and this will allow us to discuss double copy at Section \ref{doublecopy}.


\subsection{Hamiltonian approach}
\label{altH}

In this section  we present  an alternative approach to the ED lift with a special emphasis put on the Hamiltonian formalism. This is motivated by the fact that the PU model is a  bi-Hamiltonian system;  thus we cannot consider only equations of motion (in particular the geodesic ones); they can be obtained  from different Hamiltonian formalisms. To this end let us consider two  approaches. 
\par
First, we start with a pp-wave metric\footnote{We can eliminate the term with  $\omega $ by a redefinition of $K$; however, such a form will be more useful for us.} in Brinkmann coordinates $(X^i, U, V)$
\beq
\label{e5}
g=K(\bX,U)dU^2+2dUdV+d\bX^2+\omega\bX^T\epsilon d\bX dU,
\eeq
which includes both (\ref{Liftgcosc}) and (\ref{Liftgcoscunr}) as special cases ($\epsilon$ is the   antisymmetric matrix  with off-diagonal entries $\pm1$). The  Lagrangian of a relativistic particle in   an electromagnetic potential $A_\mu$ is of the form  
 \beq
 \label{r1}
 L=-m\sqrt{-g_{\mu\nu}  {{\dot X}^\mu} { \dot X}^\nu }+eA_\mu {\dot X}^\mu,
 \eeq
 where dot refers to the derivative w.r.t.  an arbitrary parameter $\tau$ and we adopt the conventions $(-,+,+,+)$.  In our further considerations  we will restrict to the form  ${\bf A}=0$,   $A=A_U(U,{\bf X})$  and  $A_V=0$ where   $U=(X^3-X^0)/\sqrt{2}$  is the ``light-cone" coordinate. 
 Then  we have
\beq
\label{r2}
L=-m\sqrt{-2 \dot U \dot V- \dot {\bf X}^2-K \dot U^2-\omega\bX^T\epsilon \dot \bX  \dot U}+eA \dot U.
\eeq
 Since $L$ is homogeneous of the  first degree in velocities  the Hamiltonian is zero. To  describe the   evolution one has to fix the gauge  by choosing  a time parameter $\tau=\tau(X^\mu)$.  
Let us  fix the gauge demanding $\tau=\tau(U)=U$.  At the Lagrangian level  this leads to 
\beq
\label{r5}
L=-m\sqrt{-2 V' - {({\bf X}')}^2-K-\omega\bX^T\epsilon  \bX' }+eA.
\eeq    
where prime  stands for  the derivative w.r.t. $U$. 
Then one  of e.q.m. gives  $-2 V' - ({\bf X}')^2-K-\omega\bX^T\epsilon  \bX' =const=m^2/ P_V^2$  where $P_V$ is the canonical momentum conjugated to $V$ coordinate; moreover, we see that  $\tau=U$  becomes   the affine parametrization.
 Now, we can  form the Hamiltonian formalism.  The  phase  space obtained is six-dimensional,  $(V,P_V)$ and  $({\bf X}, {\bf P})$ with the  canonical Poisson brackets, and the  Hamiltonian, for positive $P_V$,  is  of the following form
 \beq
 \label{r7}
 H=\frac{({\bf P}+{ \omega P_V}\epsilon \bX /2)^2+m^2}{2P_V}-eA-\frac{P_V}{2}K.
 \eeq
The above  Hamiltonian  does not depend on the $V$ coordinate and thus $P_V=const$.  In consequence for    the particle with fixed  momentum $P_V$  and $m$ the dynamics of the transversal coordinates   $({\bf X}, {\bf P})$  is governed by the Hamiltonian of the form \eqref{r7}. 
\par 
An alternative and more rigorous approach is based  on the  Dirac method for  the  time-dependent constraints (see e.g. \cite{GT}). 
 To this end let us  note that the canonical momenta for  the Lagrangian \eqref{r2}  
\beq
P_U=\frac{m \dot V+mK\dot U+m\omega\bX^T\epsilon\dot \bX/2}{\sqrt{-g_{\mu\nu}{\dot X^\mu}{\dot X^\nu}}}+eA,\quad P_V=\frac{m\dot U}{\sqrt{-g_{\mu\nu}{\dot X^\mu}{\dot X^\nu}}},\quad {\bf P}=\frac{m\dot \bX-m{\dot U}\omega\epsilon\bX/2}{\sqrt{-g_{\mu\nu}{\dot X^\mu}{\dot X^\nu}}},
\eeq
 satisfy the condition $2(-P_U+eA+\frac{K}{2}P_V)P_V-(  {\bf P}+P_V\omega\epsilon\bX/2)^2=m^2$; moreover, $\textrm{sign}(P_V)=\textrm {sign}(\dot U)$ . In consequence,  we arrive at  the constraint 
 \beq
 \phi=P_U-eA-\frac{K}{2}P_V+\frac{({\bf P} + P_V\omega\epsilon\bX/2)^2+m^2}{2P_V}.
 \eeq
Due to the above constraint we can  only express the velocities $V,\dot \bX$ in terms of momenta $P_V,{\bf P}$  (there  remains not specified   function $\dot U$).  Namely,   we obtain
 \beq
 \dot \bX =\frac{\dot U}{P_V}({\bf P} + P_V\omega\epsilon\bX/2), \quad \dot V=-\dot U\left(\frac{K}{2}+\frac{{\bf P}^2+m^2}{2P_V^2}-\frac 18 \omega^2\bX^2\right),
 \eeq
 and $U$ is an arbitrary function of $\tau$. 
 In view of the above the Hamiltonian $H_c=P_U \dot U+ P_V\dot V+{\bf P}\cdot\dot\bX-L $ takes the form 
 \beq
 H_c=\dot U \phi ,
 \eeq
 thus $\phi$ is a first class constraint. We break   the reparametrisation symmetry by adding a new constraint, 
 \beq
 \psi=U-\tau. 
 \eeq
 Then $\{\psi,\phi\}=1$ and thus $\psi$ and $\phi$  belong to  the second   class constraints. Moreover, demanding
 \beq
\psi'=\partial_\tau\psi +\{\psi,H_c\}=0,
\eeq
(i.e.  no  further constraints) we have $\dot U=1$  and $H_c=\phi$ (for $\textrm{sign}(P_V)=-1$   we should consider the constraint $ \psi=U+\tau$, then $H_c=-\phi$).   Unfortunately  the constraint $\psi$  depends on $\tau$ thus we cannot  directly  perform the Dirac procedure (it  is valid for  the time-independent case). To get rid of this problem we perform the canonical transformation   changing only the  one  variable $\tilde U=U-\tau$  ($\tilde P_\mu=P_\mu $, $\tilde V=V$, $\tilde \bX=\bX$). The  corresponding generating  function  $W$  is of the form 
\beq
W(X^\mu,\tilde P_\nu,\tau)=X^\mu \tilde P_\mu-\tau \tilde P_{\tilde U}.
\eeq
Then $\psi=0$ is equivalent to  $\tilde\psi=\tilde U =0 $ ($\tau$  independent)  
and the Hamiltonian takes the form 
\beq
\label{r8}
\tilde H= H_c+\partial_\tau W=\phi-P_U=\frac{({\bf P}+{ \omega P_V}\epsilon \bX/2 )^2+m^2}{2P_V}-eA-\frac{P_V}{2}K
\eeq
(with $U$ replaced by $\tilde U+\tau$). Now, we are in the position to perform the standard  Dirac procedure i.e. restrict dynamics to the constraint surface (eliminating $\tilde U$ and $\tilde P_{ \tilde U}=P_U$ variables).    Then the Hamiltonian  \eqref{r8}   coincides with \eqref{r7}; moreover, the Dirac bracket  between  $  \bX, V$  and  the corresponding canonical momenta take the standard form.  In the case of negative $P_V$,  after redefinition of $\tau\rightarrow -\tau$,  the Hamiltonian  $\tilde H$ can be rewritten in the form \eqref{r7}  with negative $P_V$. Finally, let us note that for the free particle we have $P_V\neq 0$  (the same holds when the $A,K$ vanish at infinites);  in our convention   the case $P_V<0$ describes particles    (antiparticles are with  $P_V>0$). Moreover, $\tilde H=-P_U$ (not $P_U$, despite the fact that $U$ takes the role of time).      
 Finally,  we can  skip  the constant term  $\frac{m^2}{2P_V}$  (using  the canonical transformation generated by $W=-\frac{m^2}{2P_V}U$).  So we can restrict ourselves to the  massless case, $m=0$; however we should keep in mind that in this case we exclude  particles with $P_V=0$ (massless particles traveling in the same direction that the wave).    
 \par
In summary,   the  transversal part of relativistic dynamics is described by  the two-dimensional   NR Hamiltonian,  where $U$ plays the role of time.  This is an alternative description of the ED lift;  namely, instead of  geodesic equations (equations of motion),  we  directly refer to  the Hamiltonian formalism   what, in turn, is relevant for the PU model.  
\par
The above results will form the starting point for our further considerations.  Namely, we will show that   for  some  plane GW  (and/or their electromagnetic counterparts)   the Hamiltonian \eqref{r8} can be related  (by a suitable canonical transformation)  to the Hamiltonian  \eqref{Hgcosc} with minus sign (and the non-zero parameter   $M$).


\subsection{BJR coordinates and Hamiltonian formalism}
\label{NHCPP}

Before we go further, let us recall that for the gravitational background  defined by \eqref{e5}  with the quadratic form $ K(U,\bX) =K_{ij}(U)X^iX^j $ ($\omega=0$ and $A_{\mu}=0 $),
 there is an alternative  coordinate system called the BJR coordinates in which the metric takes the form 
\begin{equation}
\label{BJRmetric}
ds^2 = a_{ij}(u) d\alpha^i d\alpha^j + 2 du ds.
\end{equation}
These coordinates are  more transparent for the symmetry analysis  and weak field approximation  of the relativistic particle; however, in contrast to the Brinkmann  ones they are not complete.  In this section we describe the canonical transformation which leads to the Hamiltonian in  the BJR coordinates (also when $\omega\neq 0$). This will allow us to describe the PU model in the  BJR framework. At the level of metrics, the following coordinate transformations
\begin{equation} 
\label{rotBtoBJR}
U=u, \quad X^i = O^i_j(u) \alpha^j,   \quad  V = s - \frac{1}{4} \alpha^i \alpha^j \frac{d a_{ij}}{du},    \qquad  a_{ij}= (O^T O)_{ij},   
\end{equation}
yield the  BJR metric (\ref{BJRmetric}) with the modified SL problem for $2\times 2$ $O$-matrix 
\begin{equation}
\label{e6}
O''_{ij} = \omega \epsilon_{ik} O'_{kj} + K_{ik} O_{kj}, \qquad   \big((O^T)'O\big)_{ij} = (O^TO')_{ij} - \omega (O^T\epsilon O)_{ij}.
\end{equation}
The BJR profile $a_{ij}$ is a $2\times 2$ symmetric matrix depending only on the coordinate $u$. Then, Killing vectors of (\ref{BJRmetric}) can be found easily as
\begin{equation} 
\label{BJRCarroll}
Y^v = \frac{\partial}{\partial s},   \quad  Y^c_i = \frac{\partial}{\partial \alpha^i},  \quad   Y^b_i = S^{ij}(u) \frac{\partial}{\partial \alpha^j} - \alpha^i \frac{\partial}{\partial s},
\end{equation}
cf. (\ref{CarrollrotBgen}).
In these coordinates, $s$ and $\alpha^i$-translations are symmetries and related conserved quantities are just canonical momenta $p_s$ and $p_i$.  $Y^b_i$ are called the Carroll boosts and $S^{ij}(u) = \int^u a^{ij} d\tilde{u}$ is the Souriau's matrix with $a^{ij} a_{jk} = \delta^i_k$. 
The BJR metric (\ref{BJRmetric}) is also a Bargmann manifold i.e., it is endowed with a covariantly constant null Killing vector $\partial_v$. Therefore, by  a null reduction the Hamiltonian in BJR coordinates will be of the form 
\begin{equation}
\label{HLNRBJR}
H_{BJR} = \frac{a^{ij} k_i k_j}{2p_s},
\end{equation}
moreover, we have $p_s=P_V$.
 First, let us note that  we should have $\ L_{BJR} = \frac{1}{2} p_s a_{ij}(u) (\alpha^i)' (\alpha^j)'$, $ (\alpha^i)' = \frac{d\alpha^i}{du}$ and 
\begin{equation}
\label{LrotBtoBJRNR}
L_{B} = L_{BJR}+ \frac{d F}{du} , 
\end{equation}
where the generating function $F$ is $F =  \frac{p_s}{4} \alpha^i \alpha^j a_{ij}' $; this can be confirmed  using   (\ref{rotBtoBJR}).
Rewriting (\ref{LrotBtoBJRNR})  as 
\begin{equation}
P_i (X^i)' - H_{B} = k_i (\alpha^i)' - H_{BJR} + \frac{dF}{du},
\end{equation}
where $k_i = p_s a_{ij} (\alpha^j)' $  are the  conserved canonical momenta. Next choosing $W$ as $F = W(X^i, k_j, u) - k_i \alpha^i$, we obtain the canonical  transformation of the associated Hamiltonians  
\begin{equation}
\label{canonrotBBJR}
H_{B} = H_{BJR} - \frac{\partial W(X^i, k_i, u)}{\partial u},
\end{equation}
augmented with the definitions
\begin{equation}
P_i = \frac{\partial W}{\partial X^i},   \quad   \alpha^i = \frac{\partial W}{\partial k_i},
\end{equation}
and the   generating function
\begin{equation}
\label{geneW}
W(X^i, k_j, u) = k_i O^{-1}_{ij} X^j + \frac{p_s}{4}  a'_{ij} O^{-1}_{ik} O^{-1}_{jl}X^k X^l.
\end{equation}
By direct calculations we check that  it relates the Hamiltonian \eqref{r7} (or \eqref{r8}) to  the Hamiltonian \eqref{HLNRBJR}.
Thus, we conclude that for the plane wave  metric  leading to   the PU oscillator  we can find, via a  canonical transformation (\ref{canonrotBBJR}),  the   description of the PU dynamics also in the  BJR coordinates.


\section{Circularly polarized periodic  gravitational wave}
\label{CPProt}

The simplest example whose underlying NR system is linked to the PU oscillator via the ED lift is the CPP GW \cite{Elbistan:2022umq}. First, we will use it as a testing ground  for our Hamiltonian approach. We will also discuss the BJR metric and its relation to the PU oscillator. Secondly, we will comment on the additional integrals of motion coming from the $\ell = \frac{3}{2}$ conformal NH group.


\subsection{Relation to PU oscillator}

The Brinkmann profile (\ref{Liftgcoscunr}) of CPP GW is
\beq
\label{e8}
K^{(1)}(U,\bX) =b\bX^T\left(
\begin{array}{cc}
\cos(\omega U) &\sin(\omega U)\\
\sin(\omega U) & -\cos(\omega U)
\end{array}
\right)\bX;
\eeq
with amplitude $b>0$ and frequency $\omega$. It is a vacuum solution of the Einstein equations.    
Let us  consider the $U$-dependent canonical transformation to the new coordinates $ {\bf x}, {\bf p}$  generated by the  function:
\beq
\label{e3}
W(\bX, {\bf p},U)={\bf p}^TR^{-1}(U)\bX,
\eeq
where  $R(U)$ is given by \eqref{e9}.
Then the   Hamiltonian \eqref{r8}  for the  profile $K=K^{(1)}$    transforms into the following one 
\beq
 H(  \bx,  {\bf p})= \frac{  {\bf p}^2}{2P_V}+\frac{bP_V}{2}((  x^2)^2-(  x^1)^2)+\frac{\omega}{2}(  x^2 p_1- x^1 p_2)
\eeq
so  it takes the form of \eqref{Hgcosc} with minus sign   after identification  $M=P_V$ and 
\beq
\label{CPPlambdas}
\Omega_+^2=\frac{\omega^2}{4}+ b = \lambda^2_1,   \qquad  \Omega_-^2=\frac{\omega^2}{4} - b = \lambda_2^2 .
\eeq
Thus for parameters  $\omega^2>4b$  the frequencies $\lambda_{1,2}^2$ (\ref{lambdas}) are positive and the underlying/transverse dynamics can be described by the PU model as outlined at Section \ref{trapPU}. In the other case, $\omega^2<4b$, we obtain the sum of  the ordinary and inverted harmonic oscillator.

Now, let us mention that we can pass to BJR coordinates via (\ref{rotBtoBJR}).
In the CPP GW case, $O$-matrix was found in \cite{Elbistan:2022umq} as
\begin{equation}
O = 
\begin{pmatrix}
\cos(\Omega_- u)  &  \frac{\omega}{2\Omega_+} \sin(\Omega_+ u) \\
-\frac{\omega}{2\Omega_-} \sin(\Omega_- u)  &  \cos(\Omega_+ u)   
\end{pmatrix},  
\label{Oshrunk}
\end{equation}
and the Souriau matrix is
\begin{equation}
\label{Sshrunk}
S = \frac{2}{(\Omega_+^2 - \Omega_-^2) \det{O}} 
\begin{pmatrix}
\scriptstyle \frac{\omega^2}{4 \Omega_+} \cos(\Omega_- u) \sin(\Omega_+ u) - \Omega_- \sin(\Omega_- u) \cos(\Omega_+ u) & -\frac{\omega}{2} \\
\scriptstyle  -\frac{\omega}{2}  &\scriptstyle  \Omega_+ \cos(\Omega_- u ) \sin(\Omega_+ u) - \frac{\omega^2}{4 \Omega_-} \sin(\Omega_- u) \cos(\Omega_+ u) 
\end{pmatrix}.
\end{equation}
One can explicitly write the symmetry generators (\ref{BJRCarroll}) and the associated conserved charges now. The BJR profile $a_{ij}(u)$ and/or the time dependent metric of the underlying NR system is
\begin{subequations}
\begin{align}
a_{11} &= \cos^2 \Omega_-u + \frac{\omega^2}{4\Omega_-^2} \sin^2\Omega_- u, \\
a_{12}&=  \frac{\omega}{2\Omega_+} \sin\Omega_+u \cos\Omega_-u - \frac{\omega}{2\Omega_-} \sin\Omega_-u \cos\Omega_+u  = a_{21}, \\
a_{22} &= \cos^2\Omega_+ u + \frac{\omega^2}{4 \Omega_+^2}\sin^2\Omega_+ u .
\end{align}
\end{subequations}  
Via (\ref{canonrotBBJR}), BJR Hamiltonian can be mapped to that of  the rotating Brinkmann coordinates (\ref{Hgcosc})  which is already related to the PU oscillator (\ref{PUHorigin}).


\subsection{Integrals of motion in rotating Brinkmann coordinates}

As we have seen above in the rotating coordinates $ (u,x,y,v)$ the transversal Hamiltonian is $u$-independent  thus the additional isometry is obviously $u-$translation. This is of course true for the CPP GW in which case the metric takes the form \eqref{Liftgcosc} with  	 $\Omega_\pm^2 $ given by \eqref{CPPlambdas}. We will analyse  symmetries and integrals of motion for this case  and next compare them with the ones for the PU model. 
   
 The Carroll symmetry of CPP GW is a particular case of (\ref{CarrollrotBgen}) as
\begin{subequations} 
\label{Carrollrot}
\begin{align}
Y^c_1 &= \cos\Omega_- u \frac{\partial}{\partial x} - \frac{\omega}{2 \Omega_-} \sin\Omega_- u \frac{\partial}{\partial y} - \frac{(\Omega_+^2 - \Omega_-^2)}{2\Omega_-} x\sin\Omega_- u \frac{\partial}{\partial v}, \\
Y^c_2 & = \frac{\omega}{2\Omega_+} \sin\Omega_+ u \frac{\partial}{\partial x} + \cos\Omega_+ u \frac{\partial}{\partial y} + \frac{(\Omega_+^2 - \Omega_-^2)}{2\Omega_+} y \sin \Omega_+ u  \frac{\partial}{\partial v}, \\
Y^b_1 &= - \frac{2}{(\Omega_+^2 - \Omega_-^2)} \left(  \Omega_- \sin\Omega_-u \frac{\partial}{\partial x} + \frac{\omega}{2} \cos\Omega_-u \frac{\partial}{\partial y}  \right) - x\cos\Omega_-u \frac{\partial}{\partial v}, \\
Y^b_2 &= - \frac{2}{(\Omega_+^2 - \Omega_-^2)} \left(  \frac{\omega}{2} \cos\Omega_+u \frac{\partial}{\partial x}  -\Omega_+ \sin\Omega_+u \frac{\partial}{\partial y}  \right) - y\cos\Omega_+u \frac{\partial}{\partial v},
\end{align}
\end{subequations}
where $Y^c_i$ and $Y^b_i$ denote the Carroll translations and Carroll boosts, respectively.  Their commutators with $\partial_u$ are found as
\begin{equation}
[\partial_u , Y_i^c] = \pm \frac{(\Omega_+^2 - \Omega_-^2)}{2} Y_i^b,   \quad   [\partial_u, Y^i_b] = \mp \frac{2 \Omega_\mp^2}{(\Omega_+^2 - \Omega_-^2)} Y_i^c. 
\end{equation} 

According to Noether's theorem, the substitution of basis vectors $\partial_\mu $ in (\ref{Carrollrot}) with the associated canonical momenta $p_\mu$ yields charges $Q (x, p)$ which are linear in momenta. Those charges are conserved along the geodesic motion
\begin{equation} 
\{Q, H_{geo}\} = 0, 
\end{equation}
where geodesic Hamiltonian is (\ref{Hgeo4d}). When projected, they become explicitly time dependent as $u$ plays the role of time for NR physics. For instance, Carroll translation $Y_1^c$ yields a conserved charge with explicit time-dependence 
\begin{equation}
\label{Q1c}
Q_1^c = \cos\Omega_- u p_x - \frac{\omega}{2 \Omega_-} \sin\Omega_- u p_y - \frac{(\Omega_+^2 - \Omega_-^2)}{2\Omega_-} M x\sin\Omega_- u,   \qquad   \frac{dQ_1^c}{du}= \{Q_1^c, H \} + \frac{\partial Q_1^c}{\partial u} = 0,
\end{equation}
where $H$ is the NR Hamiltonian in (\ref{symgcosc}).
After canonical transformations given at Section \ref{trapPU}, the projected Carroll vector fields (\ref{Carrollrot}) generate NH symmetry of the  PU oscillator with $p_v = M$ being the central extension, see eqn. $\#$ (4.27) in \cite{Elbistan:2022umq}.


\textbf{$\ell = \frac{3}{2}$ NH symmetry}:
On the other hand, it is known that the PU oscillator enjoys $\ell = \frac{3}{2}$ conformal NH symmetry provided that $\Omega_+ = 3 \Omega_- = 3 \Omega$ \cite{Andrzejewski:2014rza}.  In this particular case, the PU oscillator has two more symmetries, namely the dilation $D$ and special conformal transformation $K$. Now, we can go in the opposite direction and reversing the transformations at Section \ref{trapPU}, we carry those charges to the related CPP GW metric 
\begin{equation}
ds^2 = \delta_{ij} dx^i dx^j + 2 du dv + 2\sqrt{5} \Omega \epsilon_{ij} x^i dx^j du + \Omega^2 \Big(9 x^2 + y^2 \Big) du^2,
\end{equation}
with $\omega^2 = 20 \Omega^2$. A straightforward calculation leads extra charges
\begin{subequations}
\label{liftDK}
\begin{align}
D &= - \frac{1}{2\Omega }\left(A \cos{2 \Omega u} - B \sin{2\Omega u}  \right) , \\
K &= - \frac{1}{2\Omega^2}  \left( A \sin{2\Omega u}  + B \cos{2\Omega u}  + p_u \right) .
\end{align}
\end{subequations}
with 
\begin{subequations}
\begin{align}
A &= 4  \Omega^2 \left(  xy  + \frac{(\sqrt{5} +1)}{8 \Omega^2 p_v^2}  p_x p_y  + \frac{(2 + \sqrt{5})}{4 \Omega p_v} x p_x + \frac{\sqrt{5}}{4 \Omega p_v} y p_y  \right) p_v,  \\
B &= 4 \Omega^2 \left(  x^2 + \frac{(3 + 2 \sqrt{5})}{4 \Omega p_v} x p_y + \frac{(3\sqrt{5} + 5)}{16 \Omega^2 p_v^2 } p_y^2  + \frac{1}{4 \Omega p_v} y p_x  + \frac{(\sqrt{5} - 1)}{16 \Omega^2 p_v^2} p_x^2   \right) p_v .
\end{align}
\end{subequations}
We complete it  with the explicit expression of $H_{geo}$ (\ref{Hgeo4d}) 
\begin{equation}
H_{geo} = \frac{ \Big( p_i + \sqrt{5}\Omega \epsilon_{ij} x^j p_v \Big)^2  }{2} + p_u p_v - \frac{\Omega^2}{2} \Big(   9 x^2 + y^2  \Big) p_v^2,
\end{equation}
and canonical momenta
\begin{equation}
p_x  = \dot{x} - \sqrt{5} \Omega y \dot{u},   \quad    p_y = \dot{y} + \sqrt{5} \Omega x \dot{u}.
\end{equation}
The conservation of (\ref{liftDK}) can be verified via
\begin{equation}
 \{D, H_{geo} \} = 0  = \{K, H_{geo} \}.
\end{equation}
Their Poisson brackets with the conserved momentum $p_u$ yield
\begin{equation} 
\{ p_u, D\} = 2 \Omega^2 K + p_u, \quad   \{ p_u,  K  \} = - 2D.  
\end{equation}
It is straightforward to calculate the algebra of the charges $D$ and $K$ with the Carroll generators (\ref{Carrollrot}). Finally, we note that the form of (\ref{liftDK}) is similar to the distorted symmetries given in \cite{Elbistan:2020ffe, Dimakis:2022pks} and deserves a further analysis.


\section{Lukash plane wave and PU oscillator}
\label{Lukash}

As a new example, here we study the underlying NR system of the Lukash plane wave \cite{phonebook} and link it to the PU oscillator. Like CPP GW, Lukash plane wave is also a vacuum solution of Einstein's equations and it enjoys a $7$-dimensional symmetry algebra.
In complex coordinates, $1+3$ dimensional Lukash metric is given by
\begin{equation} 
\label{gLukash}
g_L = 2 dU dV + 2 d\bar{Z} dZ -2 C\ \text{Re} [ U^{2(i\kappa -1)} Z^2 ] dU^2, \quad    Z = \frac{X^1 + i X^2}{\sqrt{2}} .
\end{equation}
The coordinates are well defined either for $U>$ or $U< 0$ but singular at $U=0$. One can consider a non-negative wave amplitude, i.e.,  $C\geq 0$. Parameter $\kappa$ is the frequency of the wave. 
We will work in $U >0$ region where the Lukash metric (\ref{gLukash}) can be written in terms of the real Brinkmann coordinates 
\begin{equation}
\label{LukashU}
g_L  =  dX^i dX^i + 2 dU dV - \frac{C}{U^2} \Big( \big( (X^1)^2 - (X^2)^2 \big) \cos( 2\kappa\ln U) - 2 X^1X^2\sin(2\kappa\ln U  ) \Big) dU^2.
\end{equation}
In addition to 5-parameter Carroll isometry and the homothety $Y_h = 2V\partial_V + X^i \partial_i$, Lukash metric (\ref{LukashU}) is endowed with an extra isometry $Y_\kappa$
\begin{equation}
\label{Lukashexo}
Y_\kappa = U \partial_U - V\partial_V - \kappa \epsilon_{ij} X^i \partial_j
= U\partial_U  + \frac{1}{2} (X^i + 2\kappa \epsilon_{ij} X^j) \partial_i - \frac{1}{2} Y_h. 
\end{equation}
Thus, Lukash plane-wave is maximally symmetric.  In order to bring the  Lukash metric (\ref{LukashU}) to the form (\ref{Liftgcosc}), $Y_\kappa$ leads us to do the following transformations \cite{Zhang:2021lrw} 
\begin{equation}
\label{Lukashmt}
U = e^T, \quad X^i = e^{T/2} \xi^i ,  \quad V= V ,  \qquad   \xi^i = (\xi, \eta).
\end{equation}
In the new coordinates $(\xi, \eta, T, V)$ Lukash metric $g_L$ (\ref{LukashU}) is conformally related to another metric $g$ as
\begin{eqnarray}
\label{Lukashconfpre}
g_L &=& e^T g \\
g &=&  d\xi^i d\xi^i + \xi^i d\xi^i dT + 2 dT dV + \left(  -C \Big( (\xi^2 -\eta^2) \cos(2\kappa T) - 2\xi \eta \sin(2\kappa T)   \Big) + \frac{1}{4} \xi^i \xi^i  \right) dT^2,
\label{Lukashconf}
\end{eqnarray}
and the extra isometry $Y_\kappa$ (\ref{Lukashexo}) turns to be a combination of a $T$ translation, a rotation and homothety:
\begin{equation}
\label{Lukashexot}
Y_\kappa = \partial_T - V\partial_V + (\kappa \epsilon_{ij} \xi^j - \frac{1}{2} \xi^i ) \frac{\partial}{\partial \xi^i} = \partial_T - \kappa\epsilon_{ij} \xi^i \partial_j - \frac{1}{2} Y_h. 
\end{equation}

Now, let us make some observations: Firstly, in its new form (\ref{Lukashconfpre}), Lukash metric is no more a Bargmann manifold because of the conformal factor $e^T$. However, conformally related metrics share identical null geodesics. In order to investigate the  NR dynamics accompanying the  Lukash metric $g_L$, we can simply consider $g$ (\ref{Lukashconf}) which is a Bargmann manifold, and perform a null reduction there. A simple calculation shows that $Y_\kappa$ (\ref{Lukashexot}) of $g_L$ leads a screw isometry for the conformally related metric $g$  
\begin{equation}
\label{isoc}
Y_g = \partial_T - \kappa \epsilon_{ij} \xi^i \partial_j,
\end{equation} 
cf. (\ref{screwforgen}).  We can even redefine\footnote{$\xi^i d\xi^i dT$ part of the metric has no effect on geodesic equations and on NR dynamics as it is a total derivative.}  $V \to V - \frac{1}{4} \xi^i \xi^i$ which puts $g$ (\ref{Lukashconf}) into standard Brinkmann coordinates (\ref{Liftgcoscunr}) as
\begin{equation}
\label{Lukashc}
g =   d\xi^i d\xi^i  + 2 dT dV + \left(  -C \Big( (\xi^2 -\eta^2) \cos(2\kappa T) - 2\xi \eta \sin(2\kappa T)   \Big) + \frac{1}{4} \xi^i \xi^i  \right) dT^2.
\end{equation}
While the  Lukash metric $g_L$  is a vacuum solution, its conformal counterpart $g$ (\ref{Lukashc}) is not. In that, it also differs from the CPP GW. As in the case of CPP GW, extra isometry (\ref{isoc}) suggests us to make a rotation 
\begin{subequations}
\begin{align}
x &= \cos(\kappa T) \xi - \sin(\kappa T) \eta, \\
y &= \cos(\kappa T) \eta + \sin(\kappa T) \xi.
\end{align}
\end{subequations}
Then, renaming $T=u$, $V=v$ we end up with rotating Brinkmann coordinates as 
\begin{equation}
\label{grotf}
g = d x^2 + dy^2 + 2\kappa (y dx - x dy)du+ 2 dudv + (\Omega_+^2 x^2 + \Omega_-^2 y^2) du^2, 
\end{equation}
where
\begin{equation}
\label{omegaL}
\Omega_+^2 = \kappa^2 + \frac{1}{4} - C,    \quad   \Omega_-^2 = \kappa^2 + \frac{1}{4} + C, 
\end{equation}
cf. (\ref{Liftgcosc}).

In order to analyse  the dynamics, first, we follow  the null geodesic argument. Namely, the geodesic equations for $x$ and $y$ with $u$ being the affine parameter become 
\begin{subequations}
\label{Lukasheqns}
\begin{align}
x'' + 2\kappa y' - \Omega_+^2 x &= 0 , \\
y'' - 2\kappa x' - \Omega_-^2 y&=0 ,
\end{align}
\end{subequations}
where $\{' \}$ denotes derivation w.r.t. $u$. These equations are exactly in the form of (\ref{gcosc}). Next, we compute the diagonalization frequencies (\ref{lambdas})
\begin{equation}
\label{PUfL}
\lambda_1^2 = \kappa^2 -\frac{1}{4} + \sqrt{C^2 - \kappa^2},\qquad 
\lambda_2^2 = \kappa^2 -\frac{1}{4} - \sqrt{C^2 - \kappa^2}.
\end{equation}
 For $\kappa>\frac 12$ and $\kappa<C<\kappa^2+1/4$  the   quantities $\Omega_\pm^2$  and $\lambda^2$'s (\ref{PUfL}) are real and   positive and  thus  the transverse dynamics can be described by the PU oscillator.  However, due to the  bi-Hamiltonian aspects of the PU model we confirm it by the Hamiltonian approach.  
\par
First, we  compare \eqref{LukashU} with \eqref{e5}  and identify the profile $K$. Then,    by virtue of \eqref{r8}, we obtain the corresponding  Hamiltonian $H$. 
Next,  the following  generating function 
\beq
W(\bX,\tilde {\bf 	p},\tilde u)=\tilde {\bf p}^T\bX e^{-\tilde u/2}+\frac{P_V}{4}e^{-\tilde u}\bX^2,
\eeq
with  new  time $ \tilde u=\ln(U)$ leads to the new  coordinates $\tilde\bx,\tilde{\bf p} $ and  Hamiltonian  $\tilde H$ 
\beq
\label{e4}
\tilde  H=H\frac{d U}{d\tilde u}+\frac{\partial F}{\partial \tilde u},
\eeq
where  the right  hand side is expressed in terms of  the tilde coordinates. In our case   the Hamiltonian \eqref{r8}   transforms into the following one 
\beq
\tilde H=\frac{\tilde {\bf p}^2}{2P_V}-\frac{P_V}{8}\tilde \bx^2+\frac{P_VC}{2}\tilde \bx^T K^{(1)}(\tilde u)\bx ,
\eeq
where   $K^{(1)}$  given by  \eqref{e8} with $\omega=-2\kappa$.
Next, performing the canonical transformation  generated by \eqref{e3} we arrive at the Hamiltonian $ H$ 
\beq
  H=\frac{ {\bf p}^2}{2P_V}-\frac{P_V}{2}\left( x^2(1/4-C)+ y^2(1/4+C)\right)-\kappa ( y p_x- x  p_y).
\eeq
Now, the above Hamiltonian  takes the form of    \eqref{symgcosc} ((\ref{Hgcosc})  with minus sign)  with $\Omega_\pm$  given by  \eqref{omegaL}.  The related NR symplectic structure (\ref{symgcosc}) can be written as
\begin{eqnarray}
H_{NR} &=& \frac{\Pi_x^2}{2} + \frac{\Pi_y^2}{2}  - \frac{1}{2} (\Omega_+^2  x^2 + \Omega_-^2 y^2), \\
\sigma &=& d \Pi_x \wedge dx + d \Pi_y \wedge dy  -2\kappa dx \wedge dy , 
\end{eqnarray}
with $\Pi_x = p_x - \kappa y , \quad   \Pi_y = p_y+ \kappa x$ where NR mass $M$ which is reminiscent of conserved momentum $P_V$ can be restored trivially.  We find the chiral decomposition (\ref{cdcoef}) and  the coefficients (\ref{cdgcosc}) as
\begin{subequations}
\label{cdgcoscLU}
\begin{align}
\beta_- &= \frac{C - 2\kappa^2- \sqrt{C^2 - \kappa^2}}{2\kappa} , \quad \quad  \beta_+ = \frac{2\kappa(\kappa^2 + \frac{1}{4} - C)}{C - 2\kappa^2- \sqrt{C^2 - \kappa^2}} \\
\alpha_+&= \frac{-C - 2\kappa^2+ \sqrt{C^2 - \kappa^2}}{2\kappa} , \quad \quad \alpha_- = \frac{2\kappa(\kappa^2 + \frac{1}{4} + C)}{-C - 2\kappa^2+ \sqrt{C^2 - \kappa^2}};
\end{align}
\end{subequations}
and finally arrive at the PU Hamiltonian with frequencies \eqref{PUfL}.


\section{Complete  gravitational wave  pulse and PU model   }
\label{complete}

In this section we consider the third example of plane GWs whose transversal dynamics can be related  to the PU model.  Namely,  the  metric is  defined by  \eqref{e5}  with $\omega=0$ and the profile \cite{Andr2}
\beq 
K^{(2)}(U,\bX)=\frac{a}{(U^2+\varepsilon^2)^2}\bX ^T
\left(
\begin{array}{cc}
\cos(\phi(U)) &\sin(\phi(U))\\
\sin(\phi(U)) & -\cos(\phi(U))
\end{array}
\right)\bX,
\eeq
where 
\beq
\phi(U)=\frac{2\gamma}{\varepsilon}\tan^{-1}(U/ \varepsilon),
\eeq
and we assume  $a,\varepsilon, \gamma>0$.      In contrast to the Lukash case,  such a plane GW  is a complete  pulse and exhibits proper conformal symmetry   (besides five  Killing vectors and homothety).  
As above we consider  the Hamiltonian approach.  To this end  let us take the generating function 
\beq
W(\bX,\tilde {\bf p},\tilde u)=\frac{1}{\varepsilon}\tilde{\bf p}^T\bX \cos(\tilde u)+\frac{P_V}{2\varepsilon}\bX^2\sin(\tilde u)\cos(\tilde u),
\eeq
together with the time change $U=\varepsilon\tan(\tilde u)$.
Then the new Hamiltonian reads  $\tilde H(\tilde \bx, \tilde {\bf  p},\tilde u)=H\frac{d U}{d\tilde u}+\frac{\partial F}{\partial \tilde u}$. 
 After direct calculations we  find that the Hamiltonian \eqref{r8} (with  $K=K^{(2)}$)    transforms into 
\beq
\tilde  H=\frac{1}{2\tilde M}\tilde {\bf p}^2+\frac{\tilde M}{2}\tilde \bx^T\tilde \bx-\frac{\tilde M }{2}{\tilde \bx^T}K^{(1)}(\tilde u)\tilde \bx,
\eeq
with $\tilde M=\varepsilon P_V$ and   $K^{(1)}$  is defined by the parameters $b=a/\varepsilon^2$,  $\omega= 2\omega_0= 2\gamma/\varepsilon$.  Thus, using the $R$-transformation  we can perform the canonical transformation generated by \eqref{e3} to the new coordinates $ \bx,{\bf   p}$   (the term $\tilde \bx^T\tilde \bx$ takes the same form). At the end we arrive  at the Hamiltonian
\beq
 H= \frac{ {\bf p}^2}{2\tilde M}+\frac{\tilde M}{2}\left( x^2(1-\frac{a}{\varepsilon^2})+y^2(1+\frac{a}{\varepsilon^2})\right)+\omega_0( y p_x- x p_y).
\eeq
The above Hamiltonian is of the form \eqref{symgcosc} and thus   is separable.  Namely, for $P_V>0$  and $1-\omega_0^2<-a/\varepsilon^2$  we obtain, after direct computations,  the PU Hamiltonian   with   the frequencies  
\beq
\quad \lambda_1^2=1+\omega_0^2+\sqrt{\Delta}, \quad  \lambda_2^2=1+\omega_0^2-\sqrt{\Delta} ;
\eeq
where $\Delta=4\omega_0^2+\frac{a^2}{\varepsilon^4}$.  
\par
Of course, we may  ask  what happens for other values of parameters.  One  can check that  for $|1-\omega^2_0|<a/\varepsilon^2$  we obtain the sum of the inverted oscillator and ordinary one,    while for  $1-\omega^2_0>a/\varepsilon^2$  we obtain the sum of two oscillators.  The case   $P_V<0$ can  be analysed  by the  change $H\rightarrow -H$   and $\omega_0\rightarrow -\omega_0$.
Finally, one can easily find suitable values of parameters for which  the odd frequencies  and  additional integrals of motion appear.

\section{Ideal Penning trap and $n=3$ PU oscillator}
\label{Penning}

The Penning traps are used for mass spectroscopy and for measuring properties of charged particles (see the Nobel Lecture by Dehmelt \cite{Dehmelt} and the reviews \cite{BrownGabrielseRev, Blaum}).  The ideal Penning trap is a $3+1$ dimensional NR ion trap where charged ions are subject to a constant magnetic field in the 3rd direction $\bm{B} = B \hat{z}$ and an anisotropic but time independent quadrupole potential $V(x, y, z) = \frac{V_0}{4} (2z^2 - x^2 -y^2) $ satisfying $\bnabla^2 V(x, y, z) =0$ with constant $V_0$. It is this quadrupole potential that leads to the inverted oscillators. The Lagrangian of a single charged particle with mass M in an ideal Penning trap is 
\begin{equation} 
\label{PenningL}
L_P = \frac{1}{2} M (x'^2 + y'^2 + z'^2) + \frac{M\omega_c}{2} (x y' - y x' ) + \frac{1}{4}M \omega_z^2 ( x^2 + y^2 - 2 z^2),
\end{equation}
where we set $\omega_c = \frac{q B}{M}$ and $\omega_z = \frac{q V_0}{M}$ and $q$ is the charge of the particle.  The equations of motion are  
\begin{subequations}
\label{Penningeqns}
\begin{align}
x'' - \omega_c y' - \frac{1}{2} \omega_z^2 x &= 0, \\ 
y ''+ \omega_c x' - \frac{1}{2} \omega_z^2 y &= 0, \\
z'' + \omega_z^2 z &= 0.
\end{align}
\end{subequations}
$x$-$y$ part  of the dynamics is a special case of (\ref{gcosc}) with $\Omega_+^2 = \Omega_-^2 = \frac{1}{2} \omega_z^2$ and $\omega = \omega_c$. 
The motion in the $z-$direction is a decoupled harmonic oscillator with frequency $\omega_z$. The symplectic structure regarding (\ref{Penningeqns}) differs from (\ref{symgcosc}) only with a trivial addition of that harmonic motion.

It is known that the motions in the Penning trap can be separated into three independent parts \cite{BrownGabrielseRev, Blaum}. The first motion is the harmonic trapping one along the $z$-direction with axial frequency $\omega_z$. The second motion is called the circular cyclotron motion with the modified cyclotron frequency $\omega_+$ and the last one is the circular magnetron motion at magnetron frequency $\omega_-$ where
\begin{equation} 
\label{cyclotronmagnetron}
\omega_\pm = \frac{\omega_c \pm \sqrt{\omega_c^2 - 2 \omega_z^2}}{2}. 
\end{equation}
Typically in Penning traps $\omega_c > \omega_+ \gg \omega_z \gg \omega_-$. 
Interestingly, the energy of the magnetron motion is negative which is relevant for the buffer gas cooling technique \cite{Blaum}. And, this motivates us to study Penning trap in our PU setting outlined at Section \ref{trapPU}. 

Diagonalization frequencies  (\ref{lambdas}) are found to be
\begin{equation}
\lambda_{1,2}^2 = \frac{(\omega_c^2 - \omega_z^2) \pm \omega_c\sqrt{\omega_c^2 - 2 \omega_z^2}}{2},
\end{equation}
and we immediately realize that they are equal to the modified cyclotron and magnetron frequencies as
\begin{equation}
\lambda_1 = \omega_+,    \qquad  \lambda_2 = \omega_- .
\end{equation}
This attributes direct physical meaning to the PU oscillator frequencies $\lambda_1$ and $\lambda_2$. We may rewrite (\ref{LOcond}) in terms of $\omega_\pm, \omega_z$ as
\begin{equation}
\label{LOcondPenning}
\omega_+\omega_- = \frac{\omega_z^2}{2},    \quad  \omega_c^2 = \omega_+^2 + \omega_-^2 + \omega_z^2,
\end{equation}
thus they satisfy $\omega_+ + \omega_- = \omega_c$ and $\omega_+ - \omega_- = \sqrt{\omega_c^2 - 2 \omega_z^2}$. We note here an interesting fact that the condition for the system to be a PU oscillator (\ref{LOcondPenning}), i.e., $\omega_c^2 = \omega_+^2 + \omega_-^2 + \omega_z^2$, holds not only for the ideal trap but also for the real Penning traps suffering from imperfections \cite{BrownGabrielse}.  

Coefficients for chiral decomposition (\ref{cdcoef}) are computed as
\begin{equation}
\label{Penningchiral}
\beta_- = M \omega_+ = \alpha_-, \quad
\alpha_+ = M \omega_- = \beta_+ .
\end{equation}
In those coordinates, both the  Hamiltonian and symplectic form for  the Penning trap decompose
\begin{subequations}
\begin{align}
H_P &= \frac{M(\omega_+ - \omega_-)}{2} \left(  \omega_+ (X_-^1 X_-^1 +  X_-^2 X_-^2)  - \omega_- (X_+^2 X_+^2 +  X_+^1 X_+^1 )  \right)    + \frac{1}{2} (p_z^2 + \omega_z^2 z^2), \\
  \sigma_P &= M (\omega_+ - \omega_-)   (d X_+^1 \wedge dX_+^2 - dX_-^1 \wedge dX_-^2)  +  dp_z \wedge dz .
\end{align}
\end{subequations}
A final transformation to canonical coordinates
\begin{subequations}
\begin{align}
\big( M(\omega_+ - \omega_-) \omega_+ \big)^{1/2} X_-^1 &= - p_1, \quad
\left( \frac{M(\omega_+ - \omega_-)}{\omega_+}   \right)^{1/2} X_-^2 = x^1, \\
\big( M(\omega_+ - \omega_-)\omega_- \big)^{1/2} X_+^2 &= - p_2, \quad
\left( \frac{M(\omega_+ - \omega_-)}{\omega_-}  \right)^{1/2} X_+^1 = x^2 .  
\end{align}
\end{subequations}
provides us with the following Hamiltonian of 3 oscillators with alternating sign
\begin{equation}
\label{PenningfH} 
H_P = \frac{1}{2} \left[   \big( p_1^2 + \omega_+^2 (x^1)^2\big) -  \big(p_2^2 + \omega_-^2 (x^2)^2 \big) + \big(p_z^2 + \omega_z^2 z^2 \big) \right].   
\end{equation}
Comparison with (\ref{PUHOs}) shows that Penning trap becomes  the  $1+1$-dimensional $n=3$ (the 6th order) PU oscillator composed of separate physical motions of Penning trap with $\omega_\pm, \omega_z$. 

We can lift the Penning trap (\ref{PenningL}) and the associated plane wave is a straightforward extension of (\ref{Liftgcosc}) with the addition of $z$-motion (\ref{Penningeqns}c) and with $\omega = \omega_c$ and $\Omega_+^2 = \Omega_-^2 = \frac{\omega_z^2}{2}$.  This is a $1+4$ dimensional non-vacuum solution of Einstein's equations as noted in \cite{IONGW}. Therefore, we can derive conserved quantities for the Penning trap using the Carroll symmetries (3 translations, 3 boosts and $\partial_v$) via modifying (\ref{CarrollrotBgen}) as
\begin{subequations}
\label{CarrollPenning}
\begin{align}
Q_1^c &= -\frac{M}{2}(\omega_+ - \omega_-) ( x\sin\omega_- u + y \cos\omega_- u ) +p_x \cos\omega_- u - p_y \sin\omega_-u  , \\
Q_2^c &= -\frac{M}{2}(\omega_+ - \omega_- ) ( x\cos\omega_+ u - y \sin\omega_+ u ) + p_x \sin\omega_+u + p_y \cos\omega_+ u,  \\
Q_1^b &= - \frac{M}{2} (x \cos \omega_- u - y \sin \omega_u ) - \frac{1}{(\omega_+ - \omega_- )} (p_x \sin\omega_- u + p_y \cos\omega_- u),  \\ 
Q_2^b &= - \frac{M}{2} ( x \sin\omega_+ u + y \cos \omega_+ u) - \frac{1}{(\omega_+ - \omega_-)} (p_x \cos\omega_+u - p_y \sin\omega_+ u)\\
Q_3^c &= p_z \cos\omega_z u + M \omega_z z \sin\omega_z u, \\
Q_3^b &= \frac{p_z}{\omega_z}\sin\omega_z u - M z \cos\omega_z u.
\end{align}
\end{subequations}
They satisfy 
\begin{equation}
\frac{dQ}{du} = \{ Q, H_P \} + \frac{\partial Q}{\partial u} =0,
\end{equation}
where the conserved Hamiltonian $H_P$ can be obtained from the screw symmetry. Charges (\ref{CarrollPenning}) satisfy 3 independent Heisenberg algebras with the same central extension mass $p_v = M$ and they can be derived from the NH symmetry of the PU oscillator \cite{Andrzejewski:2014rza} as well. 

\textbf{Comments on $\ell = \frac{5}{2}$ conformal NH algebra}:
As in Section \ref{CPProt}, we may proceed the other way around.  By using the correspondence with the PU oscillator, we are immediately tempted to investigate whether a Penning trap can be a realization of  the $\ell = \frac{5}{2}$ conformal NH algebra or not. To this end, we need to impose the odd frequency condition \cite{Andrzejewski:2014rza} on the Penning trap frequencies $\omega_\pm, \omega_z$. As it is mentioned, in a typical Penning trap those frequencies follow the order $\omega_c > \omega_+ \gg\omega_z \gg\omega_-$. This physical condition forces us to set 
\begin{equation}
\label{PenningfNH}
\omega_+ = 5 \Omega,   \quad  \omega_z = 3\Omega,  \quad  \omega_- = \Omega.
\end{equation}
However, the above choice contradicts (\ref{LOcondPenning}). 
In a Penning trap, it is also possible that the magnetron frequency $\omega_-$ can be larger than $\omega_z$.  But this comes at a price because the trap is now operated close to stability point. In that case, one interchanges the values of $\omega_-$ and $\omega_z$ in (\ref{PenningfNH}) but again fails to satisfy (\ref{LOcondPenning}). 
Thus, we come to the conclusion that Penning trap is not a realization of the $\ell = \frac{5}{2}$ conformal NH algebra.


\section{Non-commuting coordinates and the PU oscillator}
\label{exotic}

From a theoretical point of view, our $2+1$ dimensional NR equations (\ref{gcosc}) provide a proper arena for exotic dynamics (see e.g., \cite{Duval:2001hu, Duval:2000xr, Horvathy:2010wv} and references therein).       
In solid state physics, non-commuting coordinates and their exotic algebra arise in anomalous Hall problem as a semiclassical effect of Berry's phase \cite{Niu}.  A particular case of (\ref{gcosc}), namely  the planar Hill's equations were already carried to the non-commutative plane where symmetries span two independent Heisenberg algebras with different central extensions \cite{Hill2,Zhang:2012cr}.
Just recently, Horvathy and his collaborators investigated $2+1$ dimensional Carrolian dynamics which admits two central charges \cite{Marsot:2022qkx}. Here, we will be dealing with its Galilean analogue and relate it to the PU oscillator explicitly.  

To postulate  non-commutativity of the  coordinates $(x, y)$, we modify the symplectic 2-form  in (\ref{symgcosc}) as
\begin{subequations}
\label{symcosctheta}
\begin{align}
\sigma\star &= \sigma + \theta d\Pi_x \wedge d \Pi_y, \\
H &=  \frac{\Pi_x^2}{2M} + \frac{\Pi_y^2}{2M} - \frac{M}{2} (\Omega_+^2 x^2 + \Omega_-^2 y^2), 
\end{align}
\end{subequations}
while keeping Hamiltonian the same. $\theta$ is a constant non-commutativity parameter. Poisson brackets can be easily obtained from the inverse of the modified symplectic matrix $\sigma\star$ (\ref{symcosctheta}a) as
\begin{equation}
\label{exocoPB}
\{\Pi_i, \Pi_j \} = \frac{\epsilon_{ij}  M\omega}{1- M\omega \theta},  \qquad    \{x^i, \Pi_j \} = \frac{\delta_{ij}}{1- M \omega \theta},   \qquad  \{x^i,  x^j\} = \frac{\epsilon_{ij} \theta}{1- M \omega \theta},
\end{equation}
where $x$ and $y$ are no more commuting. $\theta$ also modifies Hamilton's equations as
\begin{subequations}
\begin{align}
(x^i)' &= \frac{1}{1-M\omega\theta} \left( \frac{\Pi_i}{M} - \theta M \epsilon_{ij} (\Omega_+^2 x \delta_{j1}+ \Omega_-^2 y \delta_{j2})    \right), \\
(\Pi_i)' &= \frac{1}{1-M\omega\theta} \Big( \omega\epsilon_{ij}\Pi_j + M(\Omega_+^2 x \delta_{i1} + \Omega_-^2 y \delta_{i2})   \Big).
\end{align}
\end{subequations}
Poisson brackets (\ref{exocoPB}) are identical to the ones in the non-commutative Landau problem \cite{Zhang:2011zua}.  When $\theta$ is switched off, we recover (\ref{gcosc}), as expected.   

Decomposing our exotic system, we see that only the equations related to the symplectic form alter   
\begin{subequations}
\label{NCcoefcdgcosc}
\begin{align}
\beta_+ \beta_- &= M^2 \Omega_+^2,    \quad  \quad    \alpha_+ \alpha_- = M^2 \Omega_-^2, \\
\alpha_+ + \beta_-  - \theta \alpha_+ \beta_-&= M\omega,  \quad  \quad   \alpha_- + \beta_+ - \theta \alpha_- \beta_+= M \omega,  
\end{align}
\end{subequations}
cf. (\ref{coefcdgcosc}).  Assuming no relation between parameters $\omega, \Omega_\pm$ as before , we may solve for $\theta$-dependent coefficients $\alpha_\pm, \beta_\pm$\footnote{For instance, we find 
\begin{eqnarray}
\label{betam}
\beta_- &=& \frac{M (\omega^2 + \Omega_+^2 - \Omega_-^2 ) + \omega \sqrt{\Delta} - \theta M \Omega_-^2 (\theta M^2 \Omega_+^2 + \sqrt{\Delta})   }{2 (\omega - M \theta \Omega_-^2)}, \nonumber \\
\Delta &=&  \frac{M^2}{(\omega-M\theta\Omega_-^2)^2} ( \theta^2 M^2 \Omega_+^2 \Omega_-^2 + \Omega_-^2   - \Omega_+^2 - \omega^2  )^2    -   4 M^2 \Omega_+^2,  \nonumber
\end{eqnarray} 
cf. (\ref{cdgcosc}). In order to have simpler expressions, one may choose a specific parametrization such as
$\omega^2 = 2 (\Omega_+^2 + \Omega_-^2)$ \cite{Elbistan:2022umq}.  Then the remaining coefficients can be easily found.}. However, the expressions become rather long in that case. Thus we continue with symbols $\alpha_\pm, \beta_\pm$ without substituting their actual values in terms of $\omega, \Omega_\pm$.
 
Solutions of (\ref{NCcoefcdgcosc}) satisfy the following identities
\begin{subequations} 
\begin{align}
(\beta_- - \beta_+) (1- \theta \alpha_+) &= (\alpha_- - \alpha_+) (1- \theta\beta_+), \\
(\beta_- - \beta_+ ) (1-\theta\alpha_-) &= (\alpha_- - \alpha_+) (1- \theta \beta_-) , 
\end{align}
\end{subequations}
and they decompose the symplectic structure (\ref{symcosctheta}) as
\begin{subequations}
\label{symdenc}
\begin{align}
H&= \scriptstyle{\frac{M}{2} \left[  \frac{\beta_-}{M^2} (\beta_- - \beta_+) (X_-^1)^2 + \frac{(\alpha_- - \alpha_+)}{\beta_- } (1- \theta \alpha_-)(1 - \theta\beta_-) \lambda_1^2 (X_-^2)^2 - \frac{\alpha_+}{M^2}(\alpha_- - \alpha_+) (X_+^2)^2 - \frac{(\beta_- - \beta_+)}{\alpha_+} (1-\theta \beta_+) (1 - \theta \alpha_+) \lambda_2^2 (X_+^1)^2       \right] }, \\
\sigma &= (\beta_- - \beta_+) \left[ (1 - \theta\alpha_+) dX_+^1 \wedge dX_+^2 - (1- \alpha_-) dX_-^1 \wedge dX_-^2     \right],
\end{align}
\end{subequations}
where 
\begin{equation}
\label{lambda12nc}
\lambda_1^2 = \frac{\alpha_- \beta_-}{M^2 (1- \theta\alpha_-)(1 - \theta \beta_-)}, \quad   \lambda_2^2 = \frac{\alpha_+ \beta_+}{M^2 (1 - \theta\beta_+)(1- \theta \alpha_+)}
\end{equation}
Both the symplectic structure (\ref{symdenc}) and diagonalization frequencies (\ref{lambda12nc}) go back to the previous expressions (\ref{symdegeneric}) when $\theta = 0$. 

Solving Hamilton's equations, we obtain 
\begin{subequations}
\begin{align}
X_+^1 &= A \cos \lambda_2 t + B \sin \lambda_2 t ,   \qquad  \qquad \qquad \qquad  X_-^1 = C \cos \lambda_1 t + D \sin \lambda_1 t, \\
X_+^2 &=  -  \frac{\lambda_2 M (1- \theta \beta_+) }{\alpha_+} (A \sin \lambda_2 t - B \cos \lambda_2 t), \quad  X_-^2 =  - \frac{\lambda_1 M (1- \theta\beta_-)}{\alpha_-} (C \sin \lambda_1 t    - D \cos\lambda_1 t). 
\end{align}
\end{subequations}
Associated equations of motion are 
\begin{equation}  
(X_-^{1,2})'' + \lambda_1^2 X_-^{1,2} = 0, \quad  (X_+^{1,2})'' + \lambda_2^2 X_+^{1,2} = 0.
\end{equation}
Thus, we may pass to the Darboux coordinates via
\begin{subequations}
\begin{align}
p_1 & = - \sqrt{\frac{\beta_- (\beta_- - \beta_+)}{M}} X_-^1,    \qquad      x^1 = \sqrt{\frac{M (\alpha_- - \alpha_+)(1- \theta \alpha_-)(1-\theta\beta_-)}{\beta_-}} X_-^2 ,  \\
p_2 &= - \sqrt{\frac{\alpha_+ (\alpha_- - \alpha_+)}{M}} X_+^2,   \qquad x^2 = \sqrt{\frac{M(\beta_- - \beta_+)(1- \theta\beta_+)(1- \theta\alpha_+)}{\alpha_+}},   
\end{align}
\end{subequations}
so that our Hamiltonian becomes a combination of two harmonic oscillators with a minus sign between them (\ref{Hopm}) where frequencies  $\lambda_{1,2}$ are $\theta$ dependent (\ref{lambda12nc}). With the help of a final canonical transformation (\ref{scanon}) we obtain the Ostrogradski  Hamiltonian for the PU oscillator 
\begin{equation}
\label{HPUtheta}
H_{PU}^\theta = p_x v + \frac{p_v^2}{2} + \frac{\lambda_1(\theta)^2 + \lambda_2(\theta)^2}{2} v^2 - \frac{\lambda_1(\theta)^2 \lambda_2(\theta)^2}{2} x^2,
\end{equation}
$\lambda_{1,2}$ are positive for sufficiently small values of  $\theta$. 
The conserved charges  take the form 
\begin{subequations}
\label{QPUtheta}
\begin{align}
Q_1^c &= \sqrt{\frac{M(\beta_- - \beta_+)(1- \theta\beta_+)}{\alpha_+ (1- \theta \alpha_+)(\lambda_1^2 - \lambda_2^2)}} \left( (p_x + \lambda_2^2 v) \cos\lambda_2 u    - (p_v + \lambda_1^2 x)\lambda_2 \sin\lambda_2 u    \right), \\
Q_2^c &= \sqrt{\frac{(\beta_- - \beta_+)}{(\lambda_1^2 - \lambda_2^2)}} \left( \sqrt{\frac{M}{\beta_-}} \lambda_1 (p_v + \lambda_2^2 x)\cos\lambda_1 u +   \sqrt{\frac{\beta_-}{M}} \frac{\alpha_-}{M \lambda_1^2 (1-\theta \beta_-)(1-\theta\alpha_-)}(p_x + \lambda_1^2 v)\sin\lambda_1 u        \right) , \\
Q_1^b &= - \sqrt{\frac{M \alpha_+}{(\beta_- - \beta_+)(1-\theta\beta_+) (1-\theta\alpha_+)(\lambda_1^2 - \lambda_2^2)}}\left( (p_v + \lambda_1^2x)\cos\lambda_2 u  + \frac{1}{\lambda_2} (p_x + \lambda_2^2 v)\sin\lambda_2 u   \right), \\
Q_2^b &= - \sqrt{\frac{M}{(\beta_- - \beta_+)(\lambda_1^2 - \lambda_2^2)}} \left(\frac{\sqrt{\beta_-}}{\lambda_1(1-\theta\alpha_-)}(p_x +\lambda_1^2 v) \cos\lambda_1 u  - \sqrt{\frac{1}{\beta_-}} \frac{M^2 \lambda_1^2 (1-\theta\beta_-)}{\alpha_-} (p_v +\lambda_2^2 x)\sin\lambda_1 u  \right),
\end{align}
\end{subequations}
and when $\theta=0$, they are projected versions of (\ref{CarrollrotBgen}) subject to coordinate changes outlined above.  Indeed, we have 
\begin{equation}
\frac{d Q_{1,2}^{c,b}}{du} = \{ Q_{1,2}^{c,b} , H_{PU}^\theta\} + \frac{\partial Q_{1,2}^{c,b}}{\partial u} = 0,
\end{equation}
as in (\ref{Q1c}).
Their Poisson brackets read  
\begin{equation}
\label{QHPUtheta}
\{  Q_1^c, Q_1^b \} = \frac{M}{1-\theta \alpha_+},   \qquad  \{ Q_2^c,  Q_2^b\} = \frac{M}{1-\theta\alpha_-},
\end{equation}
and span the exotic NH algebra with two central extensions \cite{Duval:2001hu}. 
Obviously, there is a solution for the system (\ref{NCcoefcdgcosc}) with $\alpha_+ =\alpha_-$ that necessarily yields $\beta_- = \beta_+$. This case is excluded above as it leads to a vanishing symplectic structure (\ref{symdenc}). We emphasize that $\theta$ is a small parameter at hand and it is possible to find realizations such that $\beta_- > \beta_+$ and $\alpha_- > \alpha_+ $\footnote{For instance suitable modifications of the Penning trap with (\ref{Penningchiral}).}.
When $\theta=0$, the usual NH symmetry with one central extension $M$ is recovered. Above results are in line with \cite{Elbistan:2022umq} where $\theta=0$ and where parameters obey $\lambda_1 = \Omega_+, \quad \lambda_2 = \Omega_- , \quad  \omega^2 = 2 (\Omega_+^2 + \Omega_-^2)$, see section \ref{CPProt}.

\section{Other examples}
\label{other}

Below, we present other examples which fit in our framework.

\subsection{Double copy and electromagnetic configurations for the PU oscillator}
\label{doublecopy}
In this section, based on the classical double copy conjecture \cite{Luna:2015paa, Bahjat-Abbas:2017htu}, we will show that the PU dynamics  appears also in the motion  of the charged particle in some  electromagnetic fields.  
Namely,  in the Minkowski spacetime  and the light-cone coordinates we take the potential  $A = K(U, X^1,X^2) dU  $  where $K$  is the profile of the pp-wave metric (\ref{e5}).  Then the related field strength is
\begin{eqnarray}
\label{dcopyF}
F &=&  \frac{1}{2} \Big( \partial_\mu K \delta^U_\nu - \partial_\nu K \delta^U_\mu \Big) dX^\mu \wedge dX^\nu 
\equiv \frac{1}{2} F_{\mu\nu} dX^\mu dX^\nu . 
\end{eqnarray}
One can easily see that 
\begin{equation}
F \wedge F = 0=\star F \wedge F, 
\end{equation}
so the generated electromagnetic fields are singular\footnote{Such configurations sometimes called as null.}. Maxwell's equations are 
\begin{equation}
dF = 0, \quad   d\star F = \bnabla^2 K \ dX^1 \wedge dX^2 \wedge dU,
\end{equation} 
where $\bnabla^2$ is $2$-dimensional Laplacian. While the first set of equations are identity $d^2 =0$, the second set (here written in a compact manner) relates the original plane-wave (\ref{Liftgcoscunr}) to the Maxwell fields.   Next,  we  put $g=\eta$ (the Minkowski metric)  and    $\omega=0$ in the Lagrangian \eqref{r1}.  Then by virtue of  \eqref{r7} we obtain  that the transversal dynamics, i.e.  $X^1,X^2$ direction,  is governed  by the Hamiltonian 
\beq
 H=\frac{{\bf P}^2+m^2}{2P_V}-eA_U.
\eeq
Comparing it with \eqref{r7} we see that the above Hamiltonian is equal to the one in the gravitational case with the replacement  $K\rightarrow \frac{2eK}{P_V}$. In view of this  the transversal dynamics in such electromagnetic fields  reduce, after above identification, to the one presented for  the gravitational backgrounds.  In particular,  we can use the  examples  discussed in the above sections.  We start with  the CPP GW and its double copy, namely BB's vortex \cite{BBvort}  as pointed out by Ilderton \cite{ilderton}.  Specifically,  when it comes to the charged particle motion with charge $e$ via double copy,   diagonalization frequencies  become
\beq
 \lambda_{1,2}^2=\frac{\omega^2}{4}\pm  \frac {2eb}{P_V} ,
\eeq
and for suitable values of parameters  (cf. Section \ref{CPProt}) they are real and positive and thus   they become the frequencies of the  PU oscillator. The same concerns the electromagnetic counterparts of the Lukash  waves  after replacement  $C\rightarrow 2eC/P_V$  in the frequencies \eqref{PUfL} as well as the complete  pulse    discussed   in Section \ref{complete}.  In summary,  for all  the   electromagnetic fields constructed (via  double copy) from our gravitational examples  the transversal  dynamics, for suitable parameters, are described by the PU oscillator.  
\par
At the  end  let us consider  more general case, i.e. the metric given by \eqref{Liftgcosc} with non-vanishing $\omega$.  To make a double copy, first,  we transform it into the form \eqref{Liftgcoscunr} and  next   take the corresponding  electromagnetic potential (i.e. given  by the electromagnetic field  defined by  $A_U=K$  with \eqref{Liftgcoscunr}).	
Then, the transverse dynamics  of such a  system  is equivalent to  the gravitational one with the replacement $\omega^2\rightarrow\frac{2e\omega^2}{P_V}$ and $\Omega^2_\pm\rightarrow\frac{2e\Omega^2_\pm}{P_V}$ (for $e/P_V>0$). In view of this  the same replacement should be made in  frequencies  \eqref{lambdas},  this yields  a suitable scaling factor. In consequence,  we have a (in general  non-vacuum)   solution to the Maxwell equations  whose transverse dynamics can be modeled by the PU oscillator.


\subsection{Hill's equations}
\label{Hill}
 
In \cite{Hill}, the symmetries of Hill's equations and their relation to the Landau problem were discussed. They also studied the ED lift of the planar Hill problem. Here, we review this old problem in our framework presented at Sections \ref{trapPU} and \ref{planewaves}. 

In the center of mass frame, Hill's equations are given as 
\begin{subequations}
\label{Hillcosc}
\begin{align}
x'' - 2 \omega y' - 3\omega^2 x &= 0, \\
y'' + 2 \omega x' &= 0,
\end{align}
\end{subequations}   
and they correspond to ``zero frequency condition": one of the frequencies, say $\Omega_- =0$. Diagonalization frequencies (\ref{lambdas})  can be found easily 
\begin{equation}
\label{lambdasHill}
\lambda_1 = \omega,    \lambda_2 = 0.
\end{equation}
Solving for (\ref{coefcdgcosc}), we get 
\begin{equation}
\beta_- = 2 M \omega, \quad   \beta_+ = \frac{3}{2} M \omega, \quad \alpha_- = \frac{1}{2} M\omega, \quad \alpha_+ = 0 . 
\end{equation}
It is now straightforward decompose this system as in \cite{Hill, Hill2} with suitable modifications of Section \ref{trapPU}. Therefore, the  Hill  dynamics  can be considered as very a  special case of the PU oscillator, though strictly speaking positive frequencies are usually assumed in the latter.

When we lift Hill's equations (\ref{Hillcosc}), we obtain
\begin{equation}
\label{liftHill}
ds^2 = dx^2 + dy^2 + 2 du dv + 2\omega (x dy - y dx) + 3 \omega^2 x^2 du^2,
\end{equation}
which is a Bargmann manifold\footnote{Comparing with Lukash example (\ref{grotf}), we see that this is a particular case of it with
$
x \leftrightarrow \beta, \quad  y \leftrightarrow \alpha, \quad  \kappa = \omega,  
$
and a simple calculation yields
$$
\omega^2 = \frac{1}{2},  \quad   C = \frac{3}{4} .
$$ 
} as stated in \cite{Hill} and it is a non-vacuum solution. 
Having solved the SL problem for the relativistic plane wave (\ref{liftHill}) or finding constants of motion for the NR motion (\ref{Hillcosc}), one can readily obtain the Carroll vector fields 
\begin{subequations}
\label{HillCarroll}
\begin{align}
Y_c^1 & = \partial_x - \frac{3}{2} \omega u \partial_y + \frac{\omega}{2} (y - 3 \omega u x) \partial_v , \\
Y_c^2 & = \frac{\sin\omega u}{2} \partial_x + \cos\omega u \partial_y + \frac{\omega}{2} (x \cos\omega u + y\sin\omega u)\partial_v, \\
Y_1^b &= -\frac{2}{\omega} (\partial_y + \omega x \partial_v), \\
Y_2^b & = -\frac{2}{\omega} \cos\omega u \partial_x + \frac{4}{\omega}\sin\omega u \partial_y + 2 (x \sin \omega u - y \cos\omega u)\partial_v,   
\end{align}
\end{subequations}
which satisfy (\ref{Carrollplane}). It is an easy exercise to find the related BJR metric and the canonical transformation that leads to a new time-dependent NR Hamiltonian.


\subsection{Lagrange points and Rydberg atoms}

Equations (\ref{gcosc}) also define orbits of trojan asteroids in the Sun-Jupiter system near Lagrange points \cite{Bialynicki-Birula:1994bie}. Specifically, equations in the $x-y$ plane are 
\begin{subequations}
\begin{align}
x'' - 2\omega y' -\omega^2 (1-a) x &=0, \\
y'' + 2\omega x' -\omega^2 (1-b) y&= 0.
\end{align}
\end{subequations}
It is straightforward to derive diagonalization frequencies (\ref{lambdas}). The motion in the third direction is a harmonic one and it decouples. In \cite{Bialynicki-Birula:1994bie}, it was also argued that the motion of electrons in diatomic molecules interacting with circularly polarized electromagnetic wave is defined with similar equations.  Likewise, three dimensional motion of electrons under the effect of fields of rotating molecules are governed by the Hamiltonian in a canonical form, namely equations $\#$ (31,32) in \cite{BBrotdiatom} 
\begin{eqnarray} 
H_Q &=& \omega_+ a_+^\dagger a_+ - \omega_- a_-^\dagger a_- + \omega_z a_z^\dagger a_z + E \nonumber \\
\omega_\pm &=& \Omega \sqrt{\frac{2 - q \pm \sqrt{9 q^2 - 8q}}{2}}, \quad  \omega_z = \Omega\sqrt{q}, 
\end{eqnarray}
where $E$ is a constant. For stability, one demands $8/9 < q < 1$. In that manner, both  the systems are related to the Penning trap which can be mapped to a 6th order PU oscillator, see section \ref{Penning}.

\subsection{Gravitational trapping}
Lastly, we revisit trapping of particles via gravitational waves with angular momentum \cite{Bialynicki-Birula:2018nnk} where the motion of a test particle in the Bessel gravitational wave background was found from the geodesic deviation equations.  
Skipping the details, we present their result consisting of coupled planar equations, namely their eqn. $\#$ (14), 
\begin{subequations}
\begin{align}
x'' - \omega y' - (1/4 - \gamma)\omega^2 x   &=0 \\
y'' +\omega x' - (1/4 + \gamma)\omega^2 y &= 0,
\end{align}
\end{subequations}
where $\gamma$ is a constant. Depending on the sign of it, one can assign $\Omega_\pm$ and calculate $\lambda_{1,2}$ from (\ref{lambdas}) with $\lambda_1^2 > \lambda_2^2$. Then,  it is easy to observe that the diagonalization frequencies happen to be the same as the original frequencies 
\begin{equation}
\lambda_{1,2}^2 = \Omega_\pm^2 = (1/4 \mp \gamma)\omega^2,
\end{equation}
as in Section \ref{CPProt} where circularly polarized periodic gravitational wave was worked out. Following the same steps, one can map this system to a PU oscillator in a straightforward manner.


\section{Discussion}
\label{discussion}

Beginning with the NR planar Lagrangian (\ref{Lgcosc}), we have shown that the dynamics of several physical systems in different contexts like the  geodesic motion in certain  $1+3$ dimensional  plane-waves, NR motion of a charged ion in a Penning trap or the motion of an exotic particle in $2+1$ dimensions etc. all boil down to the  $1+1$ dimensional PU oscillator. We explicitly relate the (conformal) NH symmetries of the PU oscillator to the Carroll symmetry of plane waves and to the conserved charges of the Penning trap and the exotic particle.
For our physical examples, it was enough to consider real frequencies $\Omega_\pm$ and $\lambda_{1,2}$.  However, our computations can be extended to a wider parameter range. One challenging question for the future is how to deal with the degenerate case of the PU oscillator frequencies i.e., $\lambda_1 =\lambda_2$. See \cite{Sarioglu:2006vc} for a realization of this case.

Next, at Section 3, we discuss the ED lift of our NR planar dynamics  and write the generic $1+3$ dimensional plane wave (\ref{Liftgcosc}) whose underlying NR motion can be described by the  $n=2$ PU oscillator.  The integrability of the NR dynamics (\ref{gcosc}) allows us to solve the SL equation of the geodesic motion and find out the Carroll symmetry of that plane wave (\ref{CarrollrotBgen}). As an alternative to  null-geodesics argument, we also provide a  Hamiltonian approach because the PU oscillator is, in fact, a bi-Hamiltonian system. 

On the other hand, the ED lift for higher derivative theories  was worked out by Galajinsky and Masterov in \cite{GaMa2}. Having related its underlying NR mechanics to the PU oscillator, as a follow up problem we would like to find out the relation between our plane wave metric (\ref{Liftgcosc}) and the one presented in \cite{GaMa2}. As a warm-up exercise, at Section  \ref{NHCPP} we study the coordinate transformations between Brinkmann and BJR metrics and their underlying NR dynamics. 

Our findings lead us to the plane waves with extra symmetries. We revisit the previously studied CPP GW \cite{Elbistan:2022umq} with our  Hamiltonian approach.  We provide the Killing vectors of this plane wave and derived 2 additional charges originated from the dilation and special conformal transformation of the PU oscillator. A deeper understanding of this new charges e.g., their derivation from a Killing tensor is reserved for a future work where we aim to understand the ED lift of the conformal NH symmetry. 

At Section \ref{Lukash}, quite non-trivially we show that the underlying/transverse NR dynamics of the Lukash plane wave \cite{phonebook} can be mapped to a PU oscillator when it is of Bianchi type VI. One can explicitly write down the Carroll Killing vectors of this Lukash plane wave. It would be interesting to extend our parameter range so that we can include other Bianchi type Lukash solutions in our framework. Section \ref{complete} is devoted to another plane wave example exhibiting an extra conformal symmetry. Using our Hamiltonian approach, we find that the transverse dynamics, with a suitable parameter choice, is related to the PU oscillator.  

Having completed our discussion of plane waves, we focus on NR dynamics. Our first example is the motion of a charged ion in a $3+1$ dimensional Penning trap. We find that this motion can be mapped to $1+1$ dimensional but the 6th order PU oscillator. Quite nicely, the physical parameters defining the motion inside the trap, namely the modified cyclotron frequency $\omega_+$, the magnetron frequency $\omega_-$ and the axial frequency $\omega_z$ turn out to be actual frequencies of the PU oscillator.  The conserved charges of the motion in a Penning trap are found via Carroll symmetry of the related $1+4$ dimensional plane wave. 

It is known that the PU oscillator has different Hamiltonian descriptions. Our framework has the advantage that its symplectic form can be easily modified to obtain $2+1$ dimensional exotic dynamics (\ref{symcosctheta}). The non-commutative dynamics emerge in condensed matter systems like Hall effect because of momentum dependent semiclassical the  Berry phases \cite{Niu}. We show that our system can still be mapped to a $1+1$ dimensional PU oscillator but with $\theta$ dependent frequencies. As a result, the NH symmetry of the PU oscillator admits a second central extension. As a future project, we may search for possible condensed matter realizations of the  PU oscillator based on this new approach. It would be also interesting to apply the conformal bridge transformation \cite{Inzunza:2019sct, Alcala:2023ags} to our system and derive the corresponding theory. Concerning symmetries of the PU model and integrals of motion, another perspective for further investigations has been opened by the recent studies in optics and dark energy models  \cite{Guha:2020kmk, Comelli:2022evf}.

Lastly, we point out other examples like the motion under electromagnetic fields created via double copy of  the plane waves, Hill's problem etc.. The role of the Carroll symmetries for the motion under such electromagnetic configurations has been pointed out in \cite{ilderton, Andr1}. On the other hand, Maxwell equations in vacuum has a duality symmetry and the related conserved charge is called the optical helicity. It would be interesting to understand the meaning (if any) of the duality symmetry and its charge in the context of gravitational waves.  We also would like to find the relation between the gauge fields derived from conformally related metrics like in the Lukash plane wave case.


\section*{Acknowledgements}

We thank J. Balog, N. Dimakis, Jarah Evslin, Ilmar Gahramanov and his group, P. Horvathy, P. Kosi\'nski, M. Plyushchay, Dieter Van den Bleeken and Utku Zorba for discussions and valuable comments.  
ME is currently supported by TÜBİTAK under 2236-Co-Funded Brain Circulation Scheme2 (CoCirculation2) with project number 121C356.


\end{document}